\documentclass[useAMS,usenatbib]{mn2e}
\usepackage{graphicx}
\usepackage{multirow}
\usepackage{amsmath}
\usepackage{rotating}
\usepackage{lscape}
\usepackage{amssymb}
\usepackage[bf]{caption}
\usepackage{lmodern}

\title[Near-infrared counterparts to the GBS X-ray sources]{Near-infrared counterparts to the Galactic Bulge Survey X-ray source population}
\author[S. Greiss et al.]{S. Greiss,$^{1}$\thanks{E-mail: s.greiss@warwick.ac.uk} D. Steeghs,$^{1, 2}$ P. G. Jonker,$^{2,3,6}$ M. A. P. Torres,$^{3}$ T. J. Maccarone,$^{4}$ \newauthor R. I. Hynes,$^{5}$ C. T. Britt,$^{5}$ G. Nelemans,$^{6,7}$ B. T. G\"ansicke$^{1}$\\
$^{1}$Department of Physics, Astronomy and Astrophysics group, University of Warwick, CV4 7AL, Coventry, UK\\
$^{2}$ Harvard-Smithsonian Center for Astrophysics, 60 Garden Street, Cambridge, MA 02138, USA \\
$^{3}$ SRON, Netherlands Institute for Space Research, Sorbonnelan 2, 3584 CA, Utrecht, The Netherlands \\
$^{4}$ Department of Physics, Texas Tech University, Box 41051, Lubbock TX, 79409-1051 \\
$^{5}$ Department of Physics and Astronomy, Louisiana State University, 202 Nicholson Hall, Tower Drive, Baton Rouge, LA 70803, USA \\
$^{6}$ Department of Astrophysics/IMAPP, Radboud University Nijmegen, Heyendaalseweg 135, 6525 AJ, Nijmegen, The Netherlands \\
$^{7}$ Institute for Astronomy, KU Leuven, Celestijnenlaan 200D, 3001 Leuven, Belgium}
\begin{document}

\date{Accepted YYYY Month Date. Received YYYY Month Date; in original form YYYY Month Date}

\pagerange{\pageref{firstpage}--\pageref{lastpage}} \pubyear{YYYY}

\maketitle

\label{firstpage}

\begin{abstract}
We report on the near-infrared matches, drawn from three surveys, to the 1640 unique X-ray sources detected by {\it Chandra} in the Galactic Bulge Survey (GBS). This survey targets faint X-ray sources in the Bulge, with a particular focus on accreting compact objects.  
We present all viable counterpart candidates and associate a false alarm probability (FAP) to each near-infrared match in order to identify the most likely counterparts. The false alarm probability takes into account a statistical study involving a chance alignment test, as well as considering the positional accuracy of the individual X-ray sources. We find that although the star density in the Bulge is very high, $\sim$90\% of our sources have a false alarm probability $<$ 10\%, indicating that for most X-ray sources, viable near-infrared counterparts candidates can be identified. In addition to the FAP, we provide positional and photometric information for candidate counterparts to $\sim$95\% of the GBS X-ray sources. This information in combination with optical photometry, spectroscopy  and variability constraints will be crucial to characterize and classify secure counterparts. 
\end{abstract}

\begin{keywords}
X-rays: binaries -- stars: binaries -- near-infrared: stars.
\end{keywords}

\section{Introduction}
\label{intro}

Astrophysical X-ray sources range from extragalactic objects such as galaxy clusters and active galactic nuclei (AGN) to Galactic sources such as supernova remnants, coronally active stars, pulsars,  accreting systems containing compact objects and even some solar system bodies. The X-ray continuum observed in all these sources comes from different processes such as bremsstrahlung radiation, synchrotron radiation, blackbody radiation, inverse Compton scattering and atomic recombination.

All-sky X-ray surveys were created with NASA's first Earth-orbiting X-ray-only mission, Uhuru \citep{gianconnietal71} and have been updated since then with many various X-ray missions (e.g. HEAO: \citealt{nugentetal83}; RXTE: \citealt{levineetal96}; ROSAT: \citealt{vogesetal99, andersonetal03, andersonetal07}). Also, many all-sky X-ray monitors have been used to detect, identify and follow-up Galactic X-ray transient sources (e.g. RXTE: \citealt{oroszetal98, remillard99, rattietal12}; Swift: \citealt{zhangetal07, munozetal13}; MAXI: \citealt{kuulkersetal13}). ESA's XMM satellite also plays an important role in surveying the X-ray sky \citep{watsonetal03, watsonetal09}. The brightest X-ray point sources in Galactic environments tend to be accreting compact objects, making X-ray surveys a straight-forward method to detect them.

In our Milky Way, multi-wavelength studies of X-ray source populations have mainly been carried out in the Galactic Centre \citep{munoetal04, munoetal09, dewittetal10, mauerhanetal09} and the Galactic Plane \citep{grindlayetal05, servillatetal12, vandenbergetal12, nebotetal13} by exploiting the {\it Chandra} X-ray Observatory's excellent spatial resolution. The Centre suffers from extremely high extinction and crowding, making multi-wavelength follow-up of the X-ray sources very difficult. In most studies, it was found that a simple astrometric and photometric matching was not enough to find the true counterparts to the X-ray sources and additional photometric and spectroscopic data were required to confidently find the real matches. Moreover, the main focus so far has been on systems bright in the optical and/or NIR, making most confirmed sources giants, high-mass X-ray binaries (HMXBs) which contain early-type mass donors and cataclysmic variables (CVs). The extinction drops off rapidly away from the Galactic Centre making the follow-up study of X-ray sources considerably less challenging in the rest of the Galactic Plane and Bulge. The Galactic Bulge, also highly populated with X-ray sources due to the fact that it contains about 14\% of the mass of the Milky Way \citep{mcmillan11}, suffers from three times less extinction in E(B-V) than the Centre, making it a more practical region to study the Galactic X-ray population. Besides their detection, the identification of X-ray sources is crucial in these surveys.  With this in mind, the Galactic Bulge Survey was designed (GBS, \citealt{jonkeretal11}). \\

 In this paper, we search for, characterize and discuss the NIR counterpart candidates to the GBS X-ray sources. The NIR data were taken from the VISTA Variables in the Via Lactea Survey (VVV, \citealt{minnitietal10}), the Galactic Plane Survey (GPS, \citealt{lucasetal08}) from UKIRT Deep Sky Survey (UKIDSS) and the Two Micron All Sky Survey (2MASS, \citealt{skrutskieetal06}). In Section \ref{surveys}, we begin with a description of the different surveys, then in Section \ref{nir_cov} move on to compare all three NIR surveys in order to show how each one can be used for different purposes. Section \ref{ext} is then dedicated to constraining the extinction towards the GBS fields. Then we discuss the false alarm rate in finding the real NIR counterpart to the X-ray sources in Sections \ref{results} and \ref{discussion}.

\section{Surveys desciption}
\label{surveys}

\subsection{Galactic Bulge Survey (GBS)}

The Galactic Bulge Survey combines sensitivity for faint X-ray sources, the astrometric accuracy of the {\it Chandra} X-ray Observatory, with a complementary photometric optical $r'$, $i'$ and H$\alpha$ survey \citep{jonkeretal11}. The GBS has several goals which will mainly be accomplished with the discovery of accreting compact objects. Detecting X-ray accreting objects is necessary in order to understand binary formation and evolution \citep{jonkeretal11}. X-ray binary systems are numerous in their types such as low-mass X-ray binaries (LMXBs) which contain a neutron star or a black hole accreting matter from a low-mass companion (M$<$2M$_\odot$), ultra-compact X-ray binaries (UCXBs) which are LMXBs with orbital periods shorter than one hour. CVs, which consist of a white dwarf accreting matter from a late-type dwarf, are not usually classified as X-ray binaries even though they are binary systems and do emit X-rays. Binary systems are crucial for the determination of masses of compact objects, offering strong constraints on stellar evolution. In terms of our understanding of binary evolution, the common envelope phase is not yet well understood, therefore finding compact binary sources which have undergone one or two common-envelope phases will help us further understand that crucial evolutionary phase. The more binary systems we find in a well controlled sample, the better our constraints of binary formation and evolution will be. This can be done by comparing robust samples against predictions from population synthesis calculations. Such samples can be constructed by counting the number of sources of a given class in a well controlled area. This results in a necessary tool in the study of X-ray sources; the need to classify sources. Although the X-rays allow us to pinpoint possible accreting objects, more detailed follow-up through the detection of coincident counterparts at other wavelengths is necessary. Thus multi-wavelength studies of the GBS X-ray sources, as well as spectroscopic follow-up form a key component of our strategy (radio: \citealt{maccaroneetal12}; optical: \citealt{hynesetal12}; optical variability: \citealt{brittetal13}, spectroscopic: \citealt{rattietal13, brittetal13, torresetal13}).
\\

\begin{figure}
\includegraphics[trim= 0cm 0cm 0cm 0.cm, clip, scale=1]{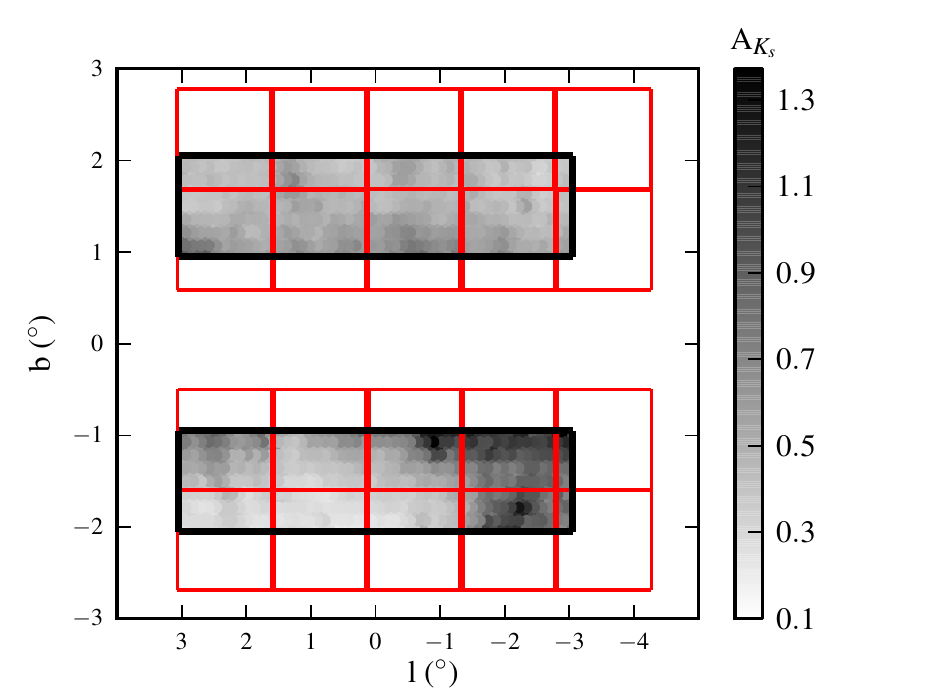}
\caption{The GBS coverage. The black boxes indicate the GBS region. In red, we show the VVV pointings which were used for the search of the NIR counterparts of the X-ray sources. The grey colour scale indicates the strength of the extinction value in the $K_s$-band (A$_{K_s}$) towards the GBS fields (see Section\,\ref{ext} for more details).\label{gbs-coverage}}
\end{figure}

The area of the sky covered in this survey is two rectangles of $l \times b=6^{\circ}\times1^{\circ}$, centred at $b = \pm 1.5^{\circ}$ (see Fig.\,\ref{gbs-coverage}). These two strips were chosen in order to avoid the Galactic Centre region ($| b |$ $<$ 1$^{\circ}$), which suffers from extremely high extinction and source confusion, while the source density is still high. The GBS is a shallow X-ray survey, of 2 ksec exposures, in order to maximize the fraction of sources that are LMXBs, while also ensuring that a large fraction of the detected sources are suitable for spectroscopic follow-up. Theoretical calculations from \citet{jonkeretal11} predict the detection of $\sim$~1600 X-ray sources in the survey region, out of which $\sim$~700 are expected to be coronally active late-type stars (single and binaries) or binary systems such as RS Canum Venaticorum (RS CVn) or W Ursae Majoris (W UMa) systems, $\sim$~600 are CVs and $\sim$~300 are LMXBs. \\

 The GBS completed the total 12 deg$^{2}$ area of the survey in both the X-ray and optical bands. Two separate X-ray energy bands (0.3-2.5 keV and 2.5-8 keV) were used to distinguish between soft and hard X-ray sources. A total of 1658 X-ray sources, with more than 3 X-ray counts, were found in the total area covered by the {\it Chandra} X-ray Observatory (see Fig.\,\ref{gbs-coverage}). \cite{jonkeretal11} published the initial list of X-ray sources detected between 2009 and 2010, containing 1234 sources. In 2011-2012, {\it Chandra} observed the remaining observations of the survey, adding another 424 X-ray sources to the list (Jonker et al. in prep). We use the source list and same naming convention as in \cite{jonkeretal11}. It is important to note that out of the initial list of published objects in \cite{jonkeretal11}, \citet{hynesetal12} found 18 duplicates, meaning that our catalogue actually contains 1640 unique X-ray sources. The main reason why duplicate sources were found in the catalogue was due to the fact that they were faint and off-axis, leading to a large PSF and poor centroiding. In our study, we will use the original catalogue of 1658 sources and comment on the duplicates in our final table containing the NIR data of their matches. \\ 
 
\subsection{The near-infrared surveys}
 
We exploit NIR data of the Bulge region in order to find the counterparts of the GBS X-ray sources. Here we present three NIR surveys which nominally cover the GBS fields: 2MASS, UKIDSS GPS and VVV. All three surveys have a different depth and coverage, each offering specific advantages in the search for the NIR counterparts of the GBS sources.
 
\subsubsection{The Two Micron All Sky Survey (2MASS)}

2MASS is a NIR survey, using $J, H$ and $K_s$ filters, which began in June 1997 and was completed in February 2001, covering 99.998\% of the celestial sphere \citep{skrutskieetal06}. It produced a Point Source Catalog containing 471 million sources and an Extended Source Catalogue of 1.7 million sources. In order to map out the entire sky, 2MASS required telescope facilities in both hemispheres. Two identical 1.3m equatorial telescopes were constructed for the survey's observations. The northern telescope is located at the Whipple Observatory at Mount Hopkins in Arizona (USA) and the southern telescope was constructed at the Cerro Tololo Inter-American Observatory at Cerro Tololo in Chile. An automated software pipeline, the 2MASS Production Pipeline System (2MAPPS), reduced each night's raw data and produced astrometrically and photometrically calibrated images and tables. The entire 2MASS data set was processed twice. The average pixel scale is 2 arcseconds per pixel. The astrometric accuracy of the 2MASS catalogue is better than 0.1 arcseconds for sources with $K_s$ $<$ 14 \citep{skrutskieetal06}.\\

This survey is reliable for sources with magnitudes up to 15.8, 15.1 and 14.3 in $J$, $H$ and $K_{s}$ respectively \citep{skrutskieetal06}, in regions which do not suffer from high densities of sources. In the Bulge, the depth is around 1.5 magnitudes shallower (see Table\,\ref{summary-table}). For this reason, we use 2MASS magnitudes solely in the case of bright sources ($K_s <$ 11.5) where the other deeper NIR surveys saturate.
  
\begin{table}
\caption{Exposure times and 5$\sigma$ limiting magnitudes in all three NIR surveys used in this paper. The GPS integration times are longer than those applied in VVV, allowing for deeper observations of the Bulge than VVV. The magnitude limits given here are for fields that are moderately crowded similar to the GBS areas.\label{summary-table}}
\begin{center}
\begin{tabular}{| c | c | c | c |}
\hline \hline
 Survey & Filters & Exposure time (s) & Depth (mag) \\ \hline
 \multicolumn{1}{ |c| }{\multirow{3}{*}{2MASS}} & \multicolumn{1}{ |c| }{$J$} & 7.8 & 14.3 \\ \cline{2-4}
 \multicolumn{1}{ |c  }{} &  \multicolumn{1}{ |c| }{$H$} & 7.8 & 13.6 \\ \cline{2-4}
 \multicolumn{1}{ |c  }{} &  \multicolumn{1}{ |c| }{$K_s$} & 7.8 & 12.8 \\ \hline
 \multicolumn{1}{ |c| }{\multirow{3}{*}{UKIDSS GPS}} & \multicolumn{1}{ |c| }{$J$} & 80 & 18.5 \\ \cline{2-4}
  \multicolumn{1}{ |c  }{} &  \multicolumn{1}{ |c| }{$H$} & 80 & 17.5 \\ \cline{2-4}
 \multicolumn{1}{ |c  }{} &  \multicolumn{1}{ |c| }{$K$} & 40 & 16.5 \\ \hline
\multicolumn{1}{ |c| }{\multirow{5}{*}{VVV}} & \multicolumn{1}{ |c| }{$Z$} & 40 & 18 \\ \cline{2-4}
  \multicolumn{1}{ |c  }{} &  \multicolumn{1}{ |c| }{$Y$} & 40 & 18 \\ \cline{2-4}
 \multicolumn{1}{ |c  }{} &  \multicolumn{1}{ |c| }{$J$} & 48 & 17 \\ \cline{2-4}
 \multicolumn{1}{ |c  }{} &  \multicolumn{1}{ |c| }{$H$} & 16 & 16.5 \\ \cline{2-4}
 \multicolumn{1}{ |c  }{} &  \multicolumn{1}{ |c| }{$K_s$} & 16 & 16 \\ \hline
\end{tabular}
\end{center}
\end{table}

\subsubsection{UKIDSS Galactic Plane Survey (GPS)}

 UKIDSS is the UKIRT (United Kingdom Infrared Telescope) Deep Sky Survey, which began in May 2005. It consists of five different surveys, each covering different areas of the sky, with the use of five near-infrared broadband filters ($ZYJHK$) as well as a narrowband one ($H_{2}$), and with a total area of 7500 deg$^{2}$ \citep{lawrenceetal07}. These surveys all use the Wide Field Camera (WFCAM), mounted on UKIRT, a  3.8 metre infrared reflecting telescope located on Mauna Kea in Hawaii. The projected pixel size is 0.4 arcseconds and the total field of view is 0.207 deg$^{2}$ per exposure. The data are reduced and calibrated at the Cambridge Astronomical Survey Unit (CASU), using a dedicated software pipeline. They are then transferred to the WFCAM Science Archive in Edinburgh\footnote{The data can be found on: http://surveys.roe.ac.uk/wsa/}. The nominal positional accuracy of UKIDSS is $\sim$ 0.1 arcseconds but this deteriorates to 0.3 arcseconds near the Bulge \citep{lucasetal08}.\\

The GPS maps the Galactic Plane in $JHK$ to a latitude of $\pm$ 5$^{\circ}$. The Galactic longitude limits are 15$^{\circ}$~$<$~$l$~$<$ 107$^{\circ}$ and 142$^{\circ}$~$<$~$l$~$<$~230$^{\circ}$. An additional narrow region, with $|b|$~$<$~2$^{\circ}$ and -2$^{\circ}$~$<$~$l$~$<$~15$^{\circ}$, will also be mapped in GPS. Thus the UKIDSS GPS overlaps fully with the GBS fields. However, coverage is not as complete as originally intended. We use data from DR8 of GPS, where the coverage in the K-band is about 65\% complete, whereas the J and H bands are still at about 35\% complete.
The total survey area of GPS is 1800 deg$^{2}$, in $JHK$ to a depth $K \sim 18$ mag \citep{lucasetal08}. This 5$\sigma$ limiting magnitude is given for non-crowded regions. In the Galactic Centre and Bulge, the depth of the survey is shallower (see Table\,\ref{summary-table}). UKIDSS GPS data saturates when the magnitudes in $JHK$ reach  $J<12.75$, $H<12.25$ and $K<11.5$ \citep{lucasetal08}. 

\subsubsection{VISTA Variables in the Via Lactea (VVV)}

VISTA (Visible and Infrared Survey Telescope for Astronomy) is a 4m class wide-field telescope, located at the Cerro Paranal Observatory in Chile. Its main purpose is to conduct large-scale surveys of the southern sky, in the NIR wavelength range. The camera mounted on VISTA is VIRCAM, which is a wide-field NIR camera with an average pixel scale of 0.34 arcseconds per pixel. The total effective field of view of the camera is 1.1 $\times$ 1.5 deg$^{2}$. The broadband filters used are $Z$, $Y$, $J$, $H$, and $K_{s}$, with bandpasses ranging from 0.8 to 2.5 $\mu$m \citep{minnitietal10, saitoetal12}.\\

VVV is a public NIR variability European Southern Observatory (ESO) survey. Its main goal is to construct the first precise 3-D map of the Galactic Bulge by using variable stars such as RR Lyrae stars and Cepheids \citep{minnitietal10, saitoetal12}, which are accurate primary distance indicators. 
The survey plan is to cover 520 deg$^{2}$ of the Galactic Bulge and an adjacent section of the mid-Plane. The Milky Way Bulge area which will be covered expands from $l$ $<$ $|10|^{\circ}$ and -10$^{\circ}$ $<$ $b$ $<$ +5$^{\circ}$, thus covering the GBS area. In our study, we use data from all five filters provided in VVV. The depth and exposure times in each band are given in Table\,\ref{summary-table}. The pipeline used to process the VVV data is based at CASU\footnote{We downloaded the VVV images and catalogues from \textrm{http://apm49.ast.cam.ac.uk/vistasp/imgquery/search}} and delivers reduced and calibrated images, as well as the aperture photometry for the VVV fields.  \\ 

We used VVV data from observations taken between March 2010 and September 2011, using version 1.1 of the photometric catalogues. The average seeing per night was typically 0.8 arcseconds \citep{saitoetal12}. We calculate the magnitudes using the following equation:
\begin{equation}
m = ZP - 2.5 \times \mathrm{log}(\frac{f}{exptime}) - apcor - percorr
\end{equation}
where $ZP$ is the zero-point magnitude of the given VVV tile as derived from standard star observations obtained during the same night, $f$ is the flux given in ADU, $exptime$ corresponds to the exposure time in the given filter, $apcor$ is the stellar aperture correction and $percor$ is the sky calibration correction. All these values are taken from the headers of the downloaded catalogues and are calculated during pipeline processing. 
The error on the magnitudes were also calculated, using the following equation: 
\begin{equation}
\sigma_m = \frac{2.5}{\ln 10} \frac{\sigma_f}{f}
\end{equation} 
where $\sigma_f$ is the error on the flux ($f$).

\begin{figure}
\hspace{-0.5cm}
\includegraphics[trim= 0cm 0cm 0cm 0cm, clip, scale=1]{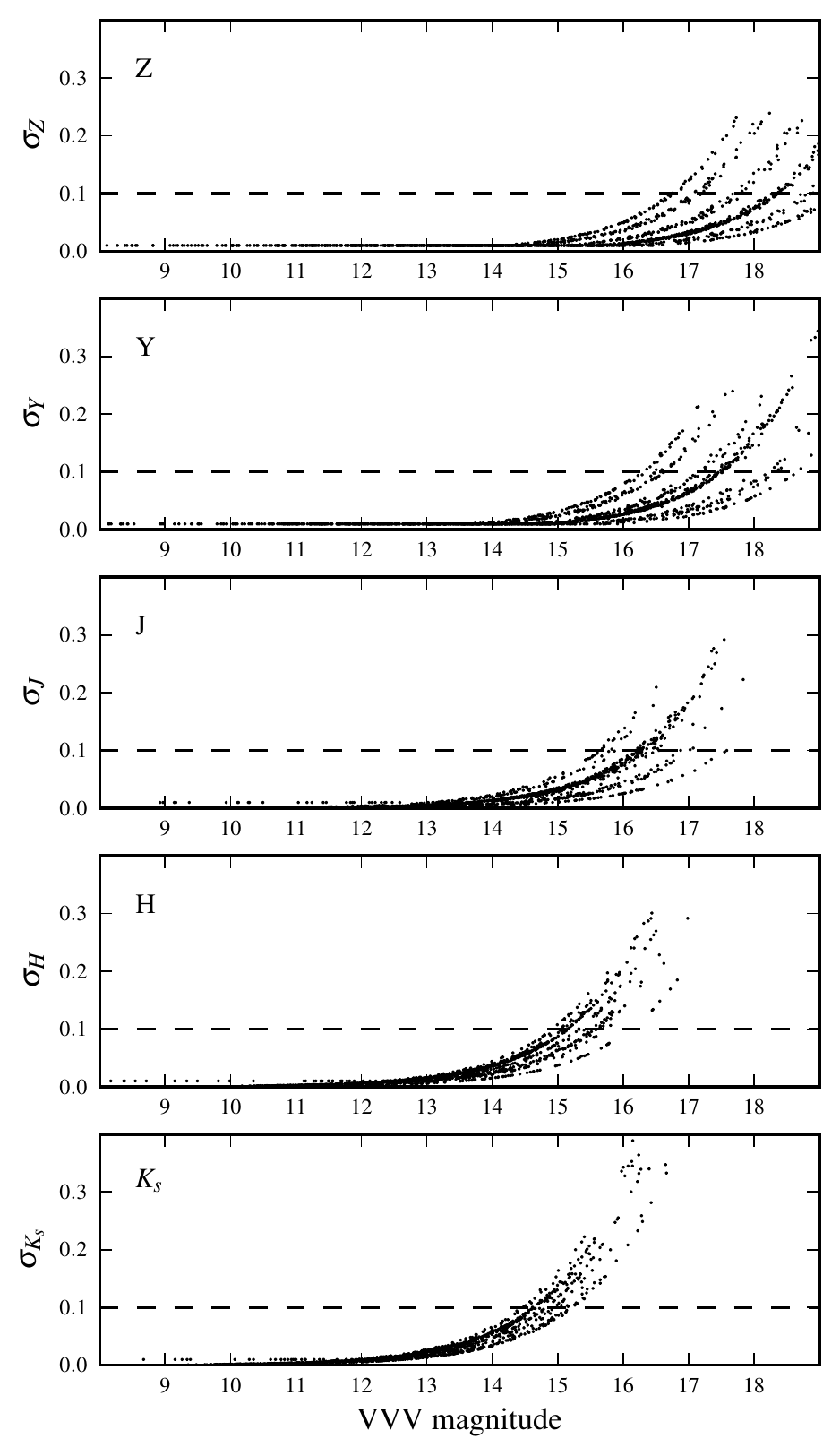}
\caption{We plot the magnitude against its uncertainty for different VVV fields. The typical 5$\sigma$ limits of sources located in the Galactic Bulge are given in Table\,\ref{summary-table}. It is clear that the different VVV fields do not have the same depth due to seeing variations from observations taken on different nights. This explains the large spread seen in the limiting magnitude values. \label{limiting-mags}}
\end{figure}

We merged all the $Z$, $Y$, $J$, $H$ and $K_s$ catalogues for each GBS source in order to work on the magnitudes and colours of any possible matches located near the X-ray positions. In order to test the quality of the photometry of the VVV data, we plot the magnitude errors against magnitudes of the nearest VVV match to the GBS sources, in all five filters (Fig.\,\ref{limiting-mags}). This gives us an indication of the limiting magnitudes of the VVV pointings we are using. Due to the dense fields of the Bulge, the actual depth is sensitive to seeing and thus covers a range around the nominal depth quoted in Table\,\ref{summary-table}. Note that VVV data saturates at $K_s \la 11.5$ mag.

\section{Near-infrared coverage of the Bulge}
\label{nir_cov}

In this Section, we compare the NIR surveys under consideration in order to show in which context each survey can be best employed. 2MASS will be useful in the case of saturated sources in VVV and UKIDSS GPS. We also show that UKIDSS GPS goes deeper than VVV, yet it does not cover the entire GBS area yet, making VVV the one with the most uniform coverage, in terms of both survey area and depth.

\subsection{Coverage}
\label{cov}

\begin{figure}
\hspace{-0.75cm}
\includegraphics[trim= 0cm 0cm 0cm 0cm, clip, scale=0.95]{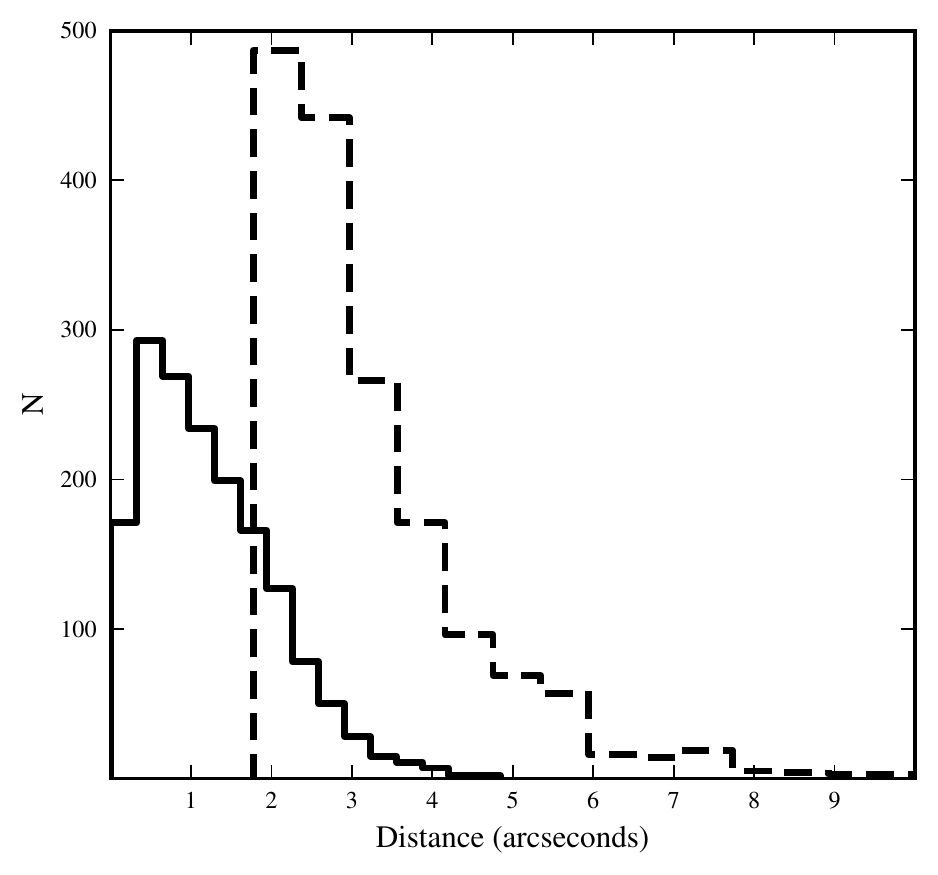}
\caption{Distribution of distances to the closest VVV matches within 5 arcseconds of the X-ray position (solid) and the 95\% confidence positional X-ray uncertainty of each GBS source (dashed). It is clear that the positional uncertainity can become very large in some cases making it impossible to choose the correct NIR match from positional coincidence alone. \label{dist-matches}}
\end{figure}

We cross-match the positions of the GBS X-ray sources with the positions of the detected stars in all three NIR surveys and compare the results we obtain from each one. To do this, we need to consider all sources of error that contribute to the absolute positional uncertainty. In Table\,\ref{coverage-table}, we give the number of matches found within respectively 5 arcseconds and 2.8 arcseconds of the X-ray position, in each filter of each survey. Those radii were chosen because 2.8 arcseconds represents the median positional uncertainty of the GBS X-ray sources corresponding to a 95\% confidence interval. The statistical component to the {\it Chandra} positional inaccuracies $P$ is calculated for each source individually using Equation 4 from \cite{evansetal10} which takes into account the number of X-ray counts and off-axis angles of each GBS source:
\begin{equation}
log~P = \begin{cases}
0.1145 \theta - 0.4957~log~C + 0.1932, & \\
 & \text{\hspace{-1.75cm} if 0.000 $<$ $log$~C $<$ 2.1393} \\
0.0968 \theta - 0.2064~log~C - 0.4260, & \\%
 & \text{\hspace{-1.75cm} if 2.1393 $<$ $log$~C $<$ 3.300} 
\end{cases}
\label{err}
\end{equation}
where $\theta$ is the off-axis angle in arcminutes and $C$ is the number of X-ray counts detected. This positional error, given in arcseconds, corresponds to a 95\% confidence interval. We mention once again that we use ACIS-I and most X-ray sources detected have a few counts (less than 10) so many {\it Chandra} positions can be uncertain by several times the size of the {\it Chandra} on-axis PSF. In addition to this component, several other terms contribute to the total positional uncertainty. One is due to some uncertainty in the spacecraft pointing, which can introduce a positional offset. The distribution of this offset reaches 0.7 arcseconds at the 95\% confidence limit\footnote{http://cxc.harvard.edu/cal/ASPECT/celmon/ - Note that we consider the ACIS-S value since it is the best determined one, with most observations taken into account for its study. This is a spacecraft correction and should not depend on the instruments on board.}. Since we cannot derive this offset from our data due to the low number of sources in one pointing, we must add a term for it.  However, it is important to note that for most sources the positional error is not dominated by this absolute correction but instead the low count rate and significant off-axis angles. Most GBS sources will have typically 5 X-ray counts and an off-axis angle of 4.5 arcminutes, leading to a positional uncertainty of 2.3 arcseconds. \citet{priminietal11} statistically characterised the positional uncertainties of {\it Chandra} Source Catalogue objects by cross-matching these to SDSS. \citet{priminietal11} found residual offsets suggesting that another component contributes to the total absolute positional accuracy. Since the radial position offset is distributed according to a Rayleigh function, the 95\% error from Eq.\ref{err} can be converted to an equivalent 1$\sigma$ error by multiplying by 0.4085 (the 95\% confidence interval for a Rayleigh PDF corresponds to a radius of 2.448\,$\times\,\sigma$). \citet{priminietal11} show that a term of 0.16" should be added in quadrature to this 1$\sigma$ error. Finally, the NIR positional uncertainty for VVV amounts to 0.08" at the 1$\sigma$ level\footnote{http://apm49.ast.cam.ac.uk/surveys-projects/vista/technical/astrometric-properties}. All these error terms are added in quadrature to obtain the total 1$\sigma$ absolute positional uncertainty $R_\mathrm{\sigma}$ that incorporates all contributions:
\begin{equation}
 R_\mathrm{\sigma}~=~\sqrt{(0.4085 \times P)^2+(0.4085 \times 0.7)^2+0.16^2+0.08^2}
\end{equation}
We show the distribution of the 95\% percentile uncertainty (R$_\mathrm{95}$) across the GBS sample in Fig.\,\ref{dist-matches}. The median value of this distribution for all GBS sources is 2.8 arcesonds. Also, more than 90\% of them have a positional uncertainty smaller than 5 arcseconds. Beyond that value, it is difficult to select the real counterpart within a 5 arcseconds radius of the X-ray position due to the high density of sources in that region (see Fig.\,\ref{dist-matches}). 

\begin{table}
\caption{Percentage of total number of valid detections found within a 5 arcseconds (upper section) radius and 2.8 arcseconds (lower section) of the X-ray positions, in 2MASS, UKIDSS GPS (DR8) and VVV \label{coverage-table}}
\begin{center}
\begin{tabular}{c c c c c c c}
\hline \hline
Survey & $Z$ & $Y$ & $J$ & $H$ & $K$ & $JHK$\\
\hline 
2MASS & - & - & 74.7 & 74.7 & 74.7 & 74.7 \\
UKIDSS GPS & - & - & 34.9 & 35.2 & 63.8 & 31.5 \\
VVV & 98.1 & 98.7 & 99.3 & 99.5 & 99.5 & 99.2 \\
\hline \hline
2MASS & - & - & 48.7 & 48.7 & 48.7 & 48.7 \\
UKIDSS GPS & - & - & 34.3 & 34.6 & 61.5 & 31.1 \\
VVV & 85.3 & 86.9 & 91.3 & 92.2 & 91.7 & 88.4 \\
\end{tabular}
\end{center}
\end{table}

We now turn to compare the NIR matches found in 2MASS and GPS with respect to the VVV matches, since the latter is the most complete survey out of the three in terms of coverage.

\subsection{VVV vs 2MASS}

 When comparing the magnitudes of the closest matches within 5 arcseconds of the X-ray positions in 2MASS and VVV,  we find a small magnitude range where both surveys are in agreement. The $J$ and $H$-bands magnitudes agree between $\sim$12 and $\sim$14$^{th}$ mag, the $K_s$-band ones between $\sim$11.5 and $\sim$13$^{th}$ mag. However, 2MASS is more reliable at the bright end, for magnitudes $<$ 12 in $J$ and $H$ and $<$ 11.5 in $K_s$, whereas VVV is more reliable in the case of fainter magnitudes. Therefore, in the case of bright sources, we use 2MASS when their NIR magnitudes are available (see Table\,\ref{coverage-table} for the number of sources with 2MASS data).

\subsection{VVV vs UKIDSS GPS}
\label{vvv-uk}

\begin{figure}
\hspace{-0.5cm}
\includegraphics[trim= 0cm 0cm 0cm 0.5cm, clip, scale=1]{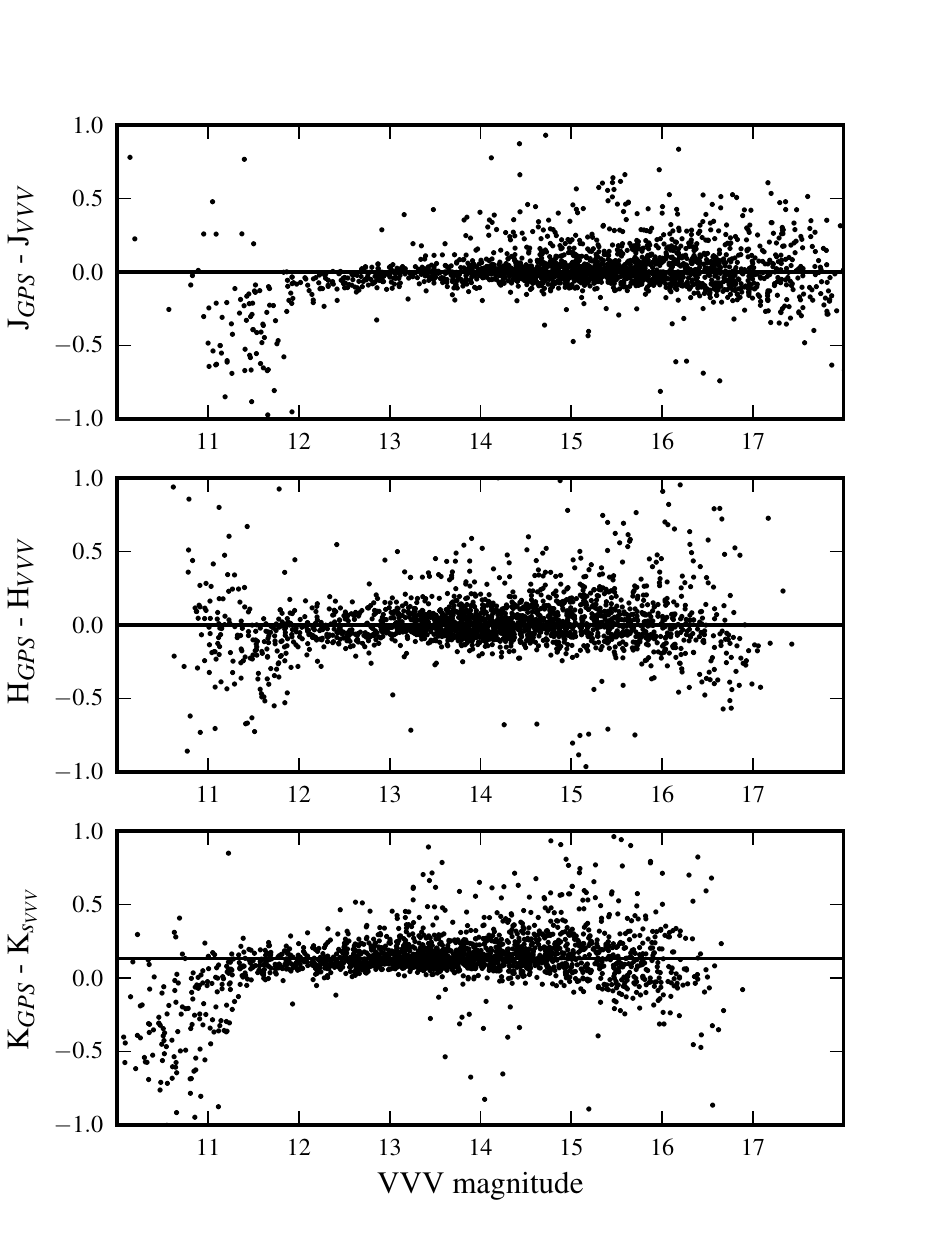}
\caption{Difference between the VVV and UKIDSS GPS magnitudes against magnitudes in $J$, $H$ and $K_s$. The solid horizontal lines correspond to the median of the difference in magnitudes between both surveys. \label{vvv-vs-ukidss}}
\end{figure}

Similarly to the work done with 2MASS, we compare the VVV matches of the GBS sources with those found in GPS. The magnitudes seem to agree in the ranges of $\sim$12 to $\sim$15 mag in all three bands (see Fig.\,\ref{vvv-vs-ukidss}). Bright sources in both surveys do not agree due to saturation problems. On the fainter end, VVV becomes less reliable and therefore starts to deviate from UKIDSS. The scattered points seen between both surveys can be explained by several reasons: many sources in the intermediate magnitude range are probably blended objects or possibly variable sources. Variable sources will be followed up in detail in a future paper. The different pipelines, filter sets and photometric systems used can also contribute towards the offsets between both surveys (clearest in the $K_s$-band), indicated with the horizontal lines in Fig.\ref{vvv-vs-ukidss}\footnote{For more information on the photometric systems, see: http://apm49.ast.cam.ac.uk/surveys-projects/vista/technical/photometric-properties}. However, GPS is not yet complete and only contains matches to $\sim$35\% of the X-ray sources in $J, H$ and $K$, whereas VVV covers over $\sim$99\% of the GBS fields. \\

\begin{figure}
\hspace{-0.5cm}
\includegraphics[trim= 0cm 0cm 0cm 0cm, clip, scale=1]{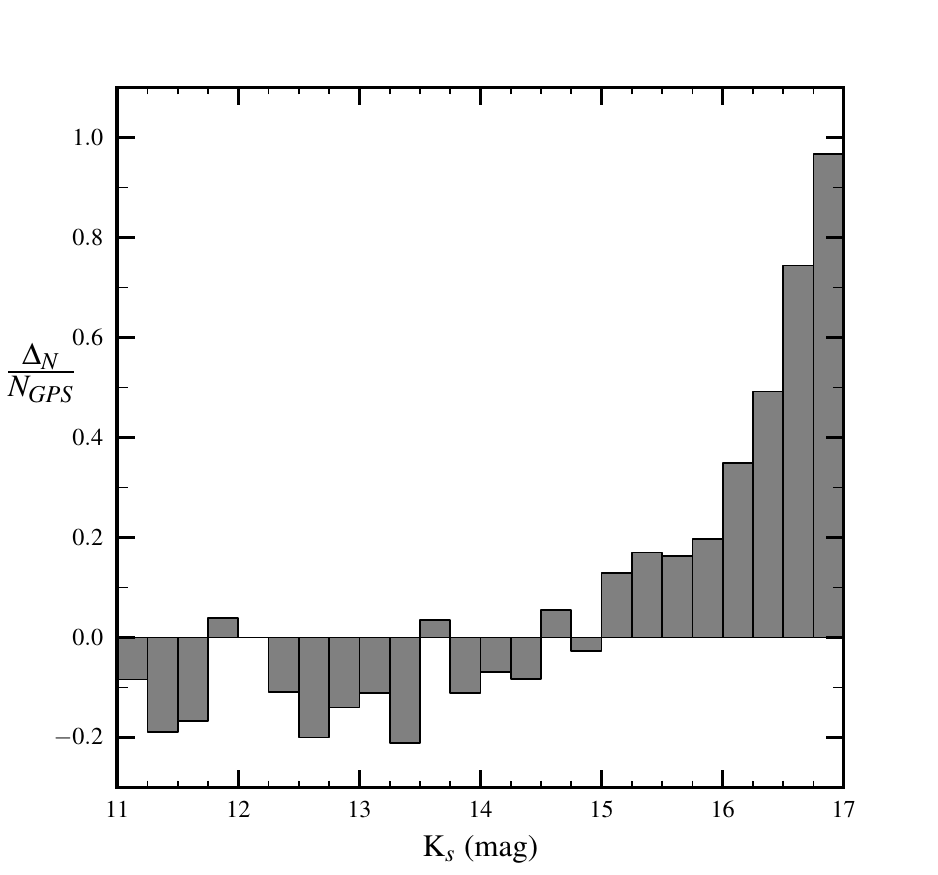}
\caption{Distribution of the fraction of detected sources UKIDSS GPS (N$_{GPS}$) and VVV (N$_{VVV}$) as a function of $K_s$ magnitude. $\Delta_N$ corresponds to (N$_{GPS}$ - N$_{VVV}$). From the increase towards 1 in the ratio towards fainter magnitudes, we conclude that the UKIDSS GPS limiting magnitude is larger than that of VVV (see text for more details). We further conclude that crowding is not a limiting factor over the whole GBS area given that the median seeing of the GPS is 1 arcseconds \citep{lucasetal08} whereas that of the VVV is 0.8 arcseconds.\label{vvv-completeness}}
\end{figure}

As seen in Table\,\ref{summary-table} GPS goes deeper than VVV. In order to confirm this statement as well as the nominal depth given in Table\,\ref{summary-table} for VVV, we show in Fig.\,\ref{vvv-completeness} the distribution of the fraction of number of sources detected in GPS and VVV, as a function of $K_s$-band magnitude. For each GBS source that has both UKIDSS GPS and VVV detections, we look for the number of GPS detections (N$_{GPS}$) and the number of VVV detections (N$_{VVV}$) within a given $K_s$-band magnitude bin. We then calculate $\Delta_N$ = N$_{GPS}$ - N$_{VVV}$, for each source, and divide by N$_{GPS}$. Finally we take the mean value of all $\frac{\Delta_N}{N_{GPS}}$ in a given magnitude bin (shown in Fig.\,\ref{vvv-completeness}). When the fraction is negative, this indicates that there are more VVV detections in the considered $K_s$-band magnitude bin. When $\Delta_N \sim 0$, both surveys are in agreement and when the fraction reaches 1, GPS dominates over VVV. We see that both surveys are on par until $K_s$ $\sim$ 16, where the VVV source catalogues become significantly incomplete, at least in the Galactic Bulge regions considered here. We further conclude that blending appears not to be the limiting factor over the whole GBS area given that the median seeing of the GPS is 1 arcsecond \citep{lucasetal08} whereas that of the VVV is 0.8 arcseconds. \\

Because we wish to have a consistent photometric system which covers a broad range of wavelengths and almost the entire solid angle of the GBS, we primarily use VVV for the search of the NIR counterparts to the X-ray sources in GBS, and use 2MASS in the case of bright matches. Our comparison shows that this gives us a secure picture of all viable counterparts down to K$\sim16$. We also report on any UKIDSS GPS detections with $K_s > $16 (see Section\,\ref{final-table}). Note that the comparison between VVV and other NIR surveys was only possible in the $JHK_s$ bands since those were the only filters in common with 2MASS and UKIDSS GPS.

\section{Exctinction}
\label{ext}

Due to the large variations on small angular scales, low spatial resolution reddening maps \citep{schlegeletal98} are not reliable in the Galactic Plane and Bulge region. We obtain the reddening values for all our GBS sources from \citet{gonzalezetal11}'s method which uses red clump (RC) giants to map the extinction towards the Bulge. They use the VVV data of the Bulge and take a 10' $\times$ 10' field, centered at ($l$ = 1.14, $b$ = -4.18) as their reference window (Baade's Window). It is an area close to the Galactic centre which suffers from relatively low amounts of reddening. The extinction towards that field is E(B-V) = 0.55 mag \citep{gonzalezetal11}. 
The relation established to obtain the extinction values is:
\begin{equation}
E(B-V) = E(B-V)_{BW} - \frac{\Delta (J - K_s)_{RC}}{(0.87 - 0.35)}
\end{equation}
where E(B-V)$_{BW}$ is the extinction towards the chosen Baade's window, and $\Delta(J - K_s)_{RC}$ is the difference between the ($J - K_s$) colour of the RC giants found in the field of unknown extinction and the ($J - K_s$) colour of the RC giants in Baade's window. The reddening values towards the GBS sources were calculated for a spatial resolution of 1.1' $\times$ 1.1' the lowest achievable with this method). The distribution of their values can be seen in Fig.\,\ref{gbs-coverage}, where we plot the extinction in the $K_s$-band (A$_{K_s}$) of the sources in Galactic coordinates. As expected, the closer the sources are to the centre, the higher the extinction is. We also notice a region of the southern strip, with $l$ $<$ -1$^{\circ}$ which suffers from the highest reddening in the GBS region.

The typical E(B-V) value towards the GBS fields is $\sim$~1.8, clearly indicating that the survey region suffers from high extinction. We note that the measured E(B-V) by \citet{gonzalezetal11} is integrated to typical distance of RC stars. Therefore, for each GBS source, the returned E(B-V) value can be lower or higher, depending on its distance.

\section{Results}
\label{results}

\begin{figure}
\hspace{-0.5cm}
\includegraphics[trim= 0cm 0cm 0cm 0cm, clip, scale=1]{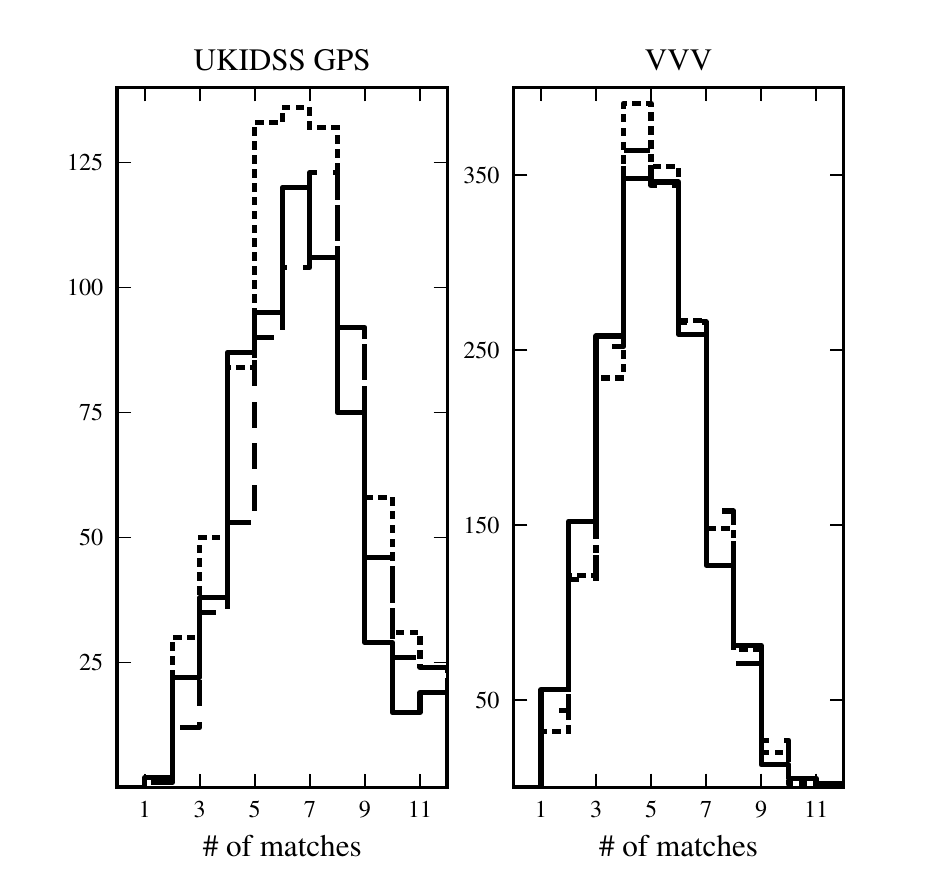}
\caption{Distribution of the number of matches found in UKIDSS GPS (left panel) and VVV (right panel) within 5 arcseconds of the X-ray position out of the total number of 1658 GBS X-ray sources. The solid line corresponds to the J-band, the dashed line to the H-band and the dotted line to the K-band. Note that the reason why the total number of sources (y-axis) in GPS is smaller than in VVV is due to the larger coverage in VVV.\label{numbers-matches}}
\end{figure}

\begin{figure*}
\hspace{-1cm}
\includegraphics[trim= 0cm 0cm 0cm 1.25cm, clip, scale=1]{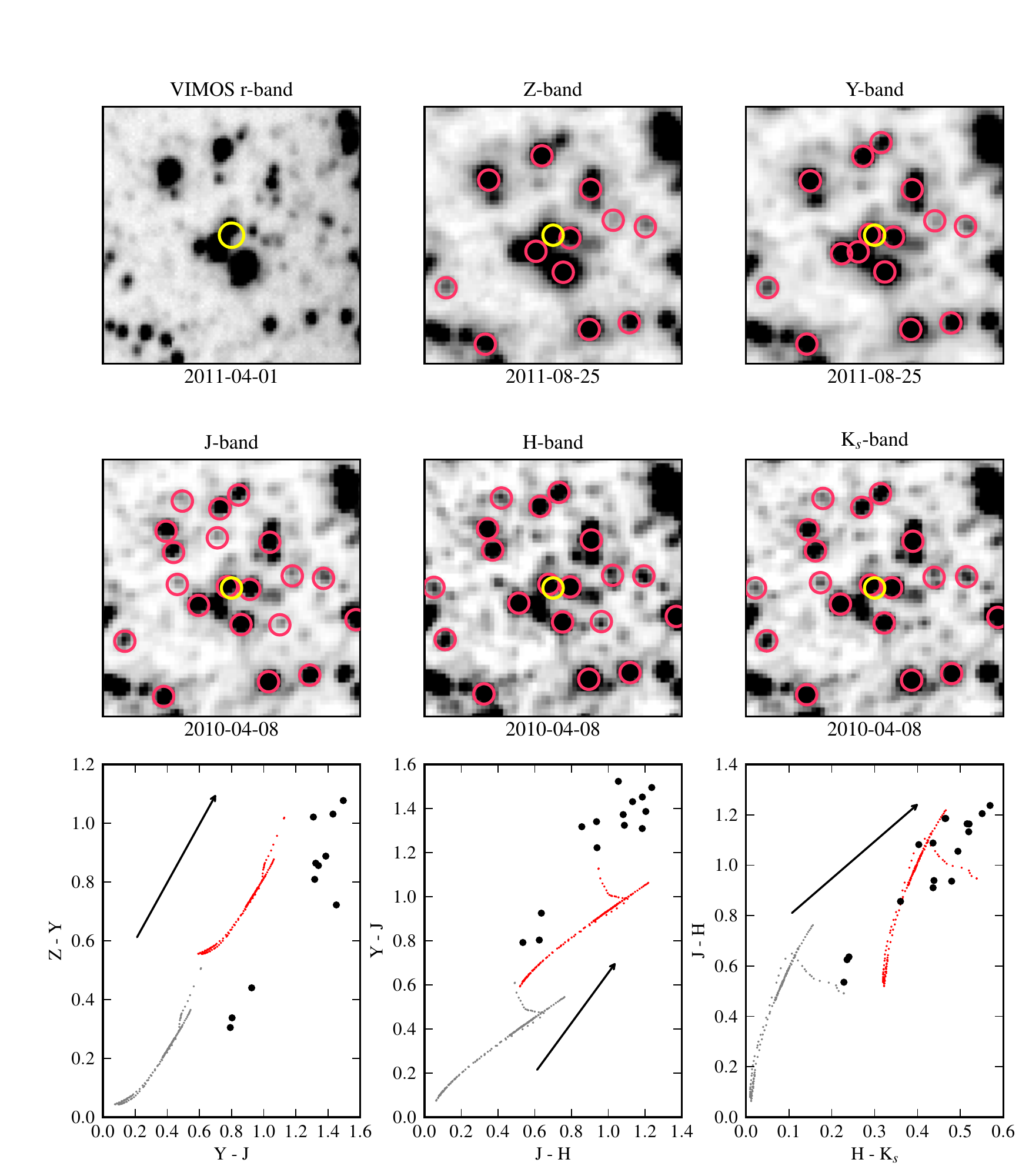}
\caption{Postage stamps of CX0377 (Wu et al. in prep), illustrating the high density of sources within 10 arcseconds of the X-ray position plotted in yellow. The red circles correspond to the VVV sources detected in each band separately. We also plot 3 colour-colour diagrams ($Z-Y$ vs $Y-J$, $Y-J$ vs $J-H$, $J-H$ vs $H-K_s$) with the VVV matches found in each case. We add reddened and un-reddened synthetic tracks of main-sequence stars, in red and grey respectively, as well as a reddening vector (with E(B-V) = 1.53). The high number of possible matches is due to the very large uncertainties in the X-ray position. Many sources suffer from blending and they only become clearer in the $K_s$-band (the seeing gets better in longer wavelengths). We also notice the non-detection of some objects in the given filters, despite their clear presence in the images. This is probably due to issues with the crowding and sky subtraction in the pipepline.  \label{postage-stamp}}
\end{figure*}

For 99.6\% of the GBS sources, we now have NIR data from VVV in $Z, Y, J, H$ and $K_s$. When data was available, we created a small catalogue with all the NIR objects found within 10 arcseconds of the X-ray position. Most sources returned over 10 neighbours, which clearly reflects on the multiple possible matches found for each X-ray source (see Fig.\,\ref{numbers-matches}-right). We also have an approximate reddening value for most GBS sources (see Fig.\,\ref{gbs-coverage}), as well as their X-ray properties from \citet{jonkeretal11}. For each {\it Chandra} source, we created a postage stamp with five VVV finder charts (10 x 10 arcseconds$^2$), in each filter, as well as three colour-colour diagrams: ($Z-Y$,~$Y-J$), ($Y-J$,~$J-H$) and ($J-H$,~$H-K_s$) (see Fig.\,\ref{postage-stamp}). We included $ZYJHK_s$ isochrones\footnote{The values were given by Stefano Rubele and Leo Girardi, members of the VVV team} in the VISTA photometric system, in order to know where the un-reddened main-sequence stars lie. These colour-colour diagrams and postage stamps were used by us to identify possible targets for spectroscopic follow-up in order to classify likely counterpart candidates. More details on the spectroscopic component of the GBS is given in \cite{torresetal13}. While these individual data sheets offer detailed insights into the specific environments around our X-ray sources, we now turn to a more robust statistical study of the counterpart candidates detected in the NIR in order to identify those that may be considered genuine NIR matches.

\subsection{Quantifying the false alarm rate}

\begin{figure*}
\begin{minipage}[t]{.475\linewidth}
\centering
\includegraphics[trim= 0cm 0cm 0cm 0.75cm, clip, scale=0.95]{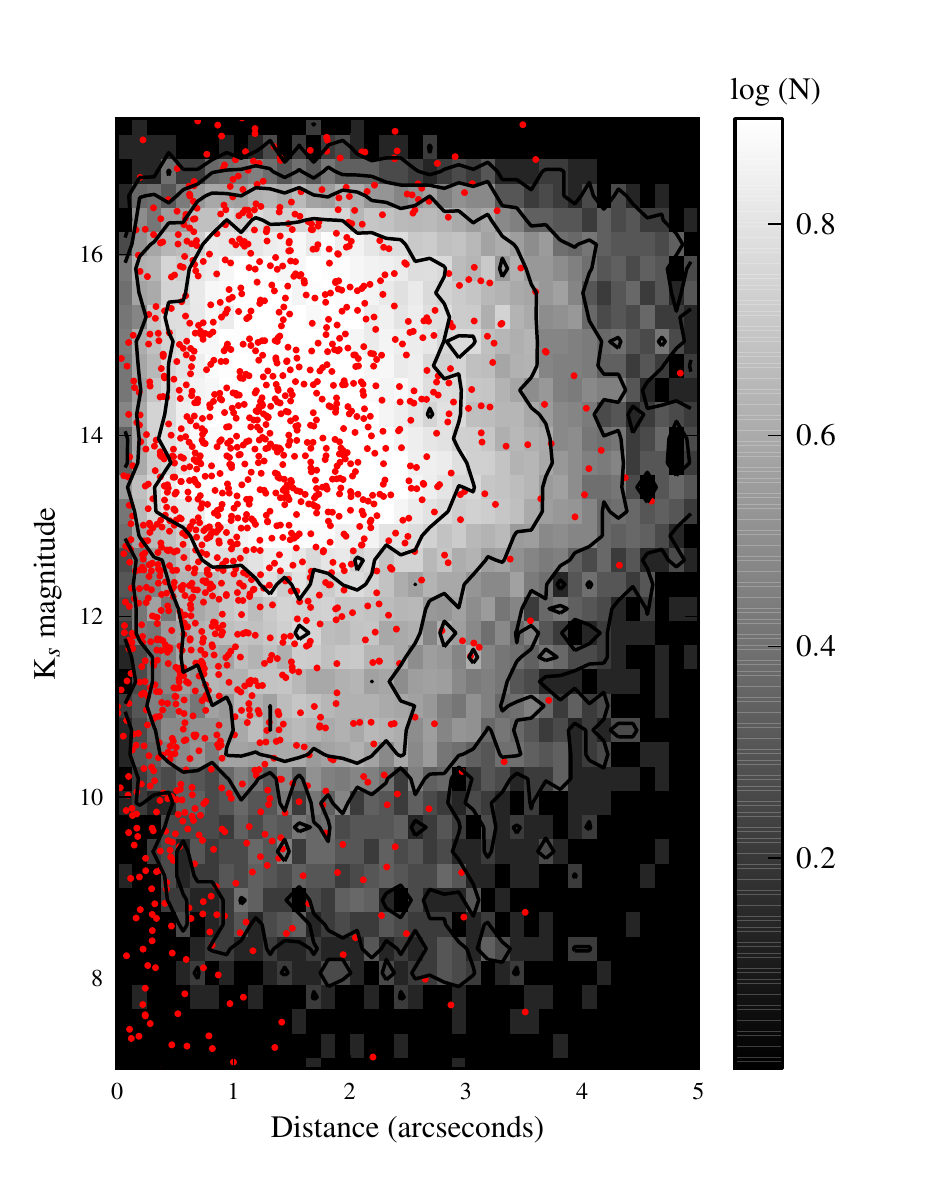}
\caption{Density plot of the $K_s$-band magnitudes of the nearest VVV matches of $\sim$ 40,000 generated sources in the Bulge against their distances to the corresponding sources. The grey scale is a normalized logarithmic scale. The red dots correspond to the nearest VVV matches to the GBS sources. Sources brighter than 8$^{th}$ magnitude are not included in this figure because they are the main focus of the study carried out by \citet{hynesetal12}. \label{density-plot-mag-vs-dist}}
\end{minipage}
\hspace{0.5cm}
\begin{minipage}[t]{.475\linewidth}
\centering
\includegraphics[trim= 0cm 0cm 0cm 0.75cm, clip, scale=0.95]{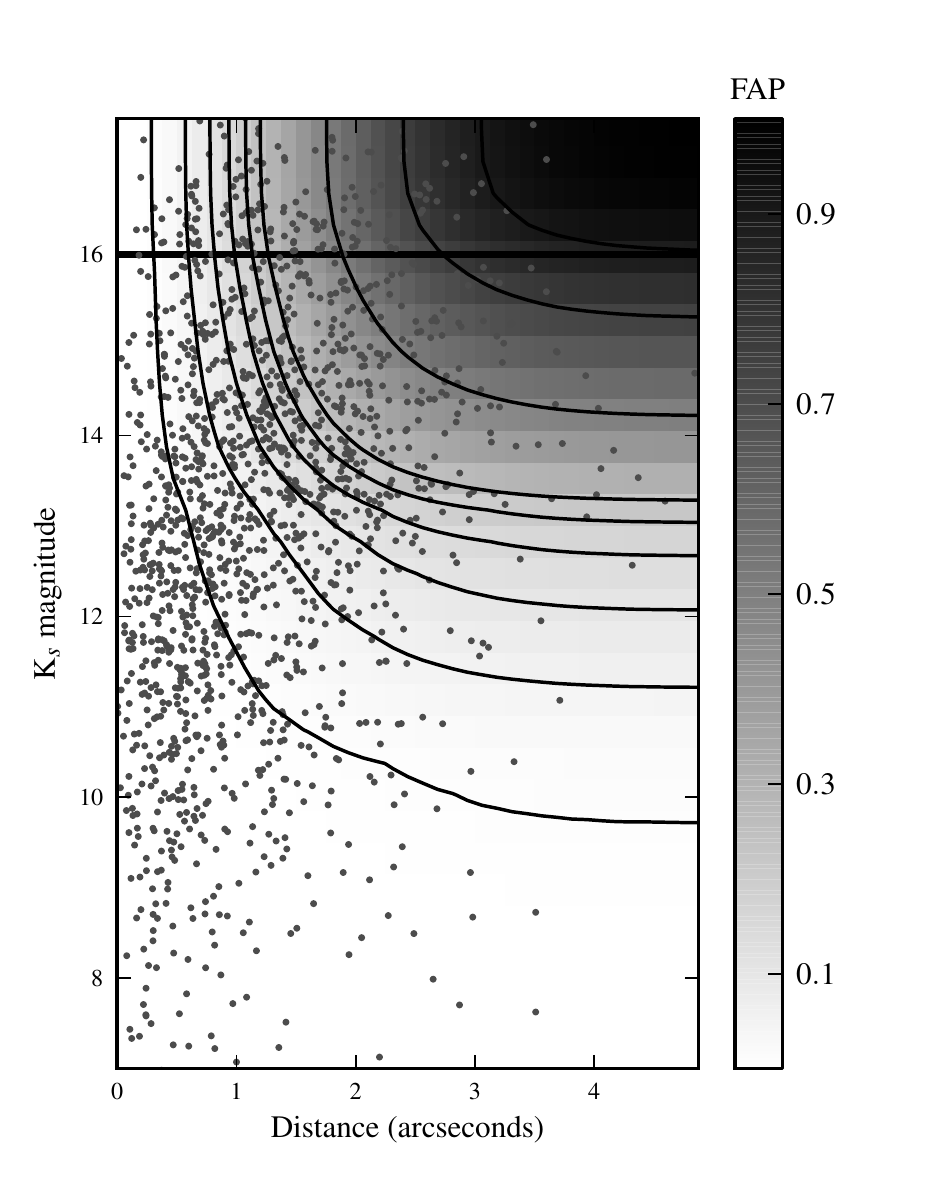}
\caption{Cumulative distribution of the FAP of having the real VVV match. The contours indicate a false alarm probabilities at 0.01, 0.05, 0.1, 0.15, 0.25, 0.5, 0.75 and 0.9. The grey dots correspond to the nearest VVV counterparts of the GBS sources. The FAP distribution in the region above the black horizontal line is artifically low, due to the lack of detected sources.\label{cum-dist}}
\end{minipage}
\end{figure*}

The goal of this study is to quantify the false alarm rate when matching the GBS X-ray sources with NIR surveys of the Bulge given the large stellar densities. Not only do we take into account the positional uncertainties of each GBS source, we also calculate a statistical false alarm probability (FAP) based on the brightness of the NIR match as well as its distance from the X-ray position. A final test is done taking into account the fact that for a given GBS source, more than one match is often detected, thus a FAP is evaluated for each match individually.

\subsection{Random matching}

In order to quantify the false alarm rate of our VVV matches, we generate a catalogue containing $\sim$40,000 random positions near the GBS source positions. In order to avoid duplicates, the sources in the generated catalogues are at least 10 arcseconds away from each other and fall in regions with 0.5$^{\circ} < |b| <$ 3$^{\circ}$ and -3$^{\circ} <$ l $<$ 4$^{\circ}$. We cross-match those random positions with the positions of stars detected in VVV and search for their nearest NIR counterparts. Such a random test preserves the specific environments our GBS sources are detected in, and also carries with it any source detection biases the survey may have. In this way it self-calibrates and is preferred over analytic estimates based on stellar densities. 
In Fig.\,\ref{density-plot-mag-vs-dist}, we show a density map (in $K_s$ mag vs distance) of the background field population of sources which could lead to false matches, and overplot in red dots the $K_s$-band magnitudes of each VVV closest match to the GBS sources against their separation from them. We notice that the counterparts of the GBS sources do not follow the same distribution as the generated sources, where the bulk of random sources fall within a defined region in the figure. This is an indication that in many cases (e.g. sources with $K_s < $ 12), the VVV matches of the GBS are not random sources.\\

We continued quantifying the false alarm rate by using this density map to create a cumulative FAP map as a function of the source's $K_s$-band magnitude and distance to it (see Fig.\,\ref{cum-dist}). The grey dots in Figure\,\ref{cum-dist} correspond to the GBS counterparts. It is important to remember that objects fainter than $K_s \sim$16 were not detected reliably, even if there is evidence for those sources in the VVV images (see Section.\,\ref{vvv-uk}). Therefore the FAP distribution in the region above the black horizontal line in Fig.\,\ref{cum-dist} is artifically low, due to the lack of detected sources. Only a handful of sources have matches fainter than $K_s \sim$ 16, but in this regime we extrapolate our FAP distribution for the GPS counterparts. Note that this analysis can be done in any filter and we illustrate it here in the $K_s$-band since it has the best coverage and suffers from lower extinction and thus tends to have the highest source densities.\\

With such a cumulative distribution at hand, we calculate the FAP ($FAP_\mathrm{random}$) for each GBS counterpart by interpolating across both source magnitude and separation to find the FAP value for that counterpart. Therefore, each NIR match within R$_\mathrm{95}$ of a GBS source will have an associated FAP, which depends on its magnitude and distance to its NIR match and reflects the density of field sources near GBS sources. The $FAP_\mathrm{random}$ obtained this way does not yet take into account the fact that multiple matches present themselves per source, nor the positional uncertainties of each GBS source. Therefore we must calculate additional false alarm probabilities based on those two criteria.

\subsection{Positional uncertainties}

In order to determine the final FAP for each NIR counterpart to the GBS sources, we need to take into account the total positional uncertainties of each GBS source as these vary considerably from source to source (see Section \ref{cov}). In each case, we calculate the cumulative density function (CDF) of the positional error, assuming it has a Rayleigh distribution: 
\begin{equation}
\frac{1}{\sigma^2} \int_0^{R_{\sigma}} \mathrm{e}^{-\frac{r^2}{2\sigma^2}}\,r dr
\end{equation}
with $\sigma$ = 1, $R_{\sigma}$ the 1$\sigma$ positional error and $r$~=~$\frac{d}{R_\mathrm{\sigma}}$, $d$~=~distance to NIR match. The FAP based on the position of the NIR match with regards to the positional uncertainty of each source is $FAP_\mathrm{position}$~=~CDF(r). 

\subsection{Total FAP}

\begin{figure}
\hspace{-0.5cm}
\includegraphics[trim= 0cm 0cm 0cm 0cm, clip, scale=0.95]{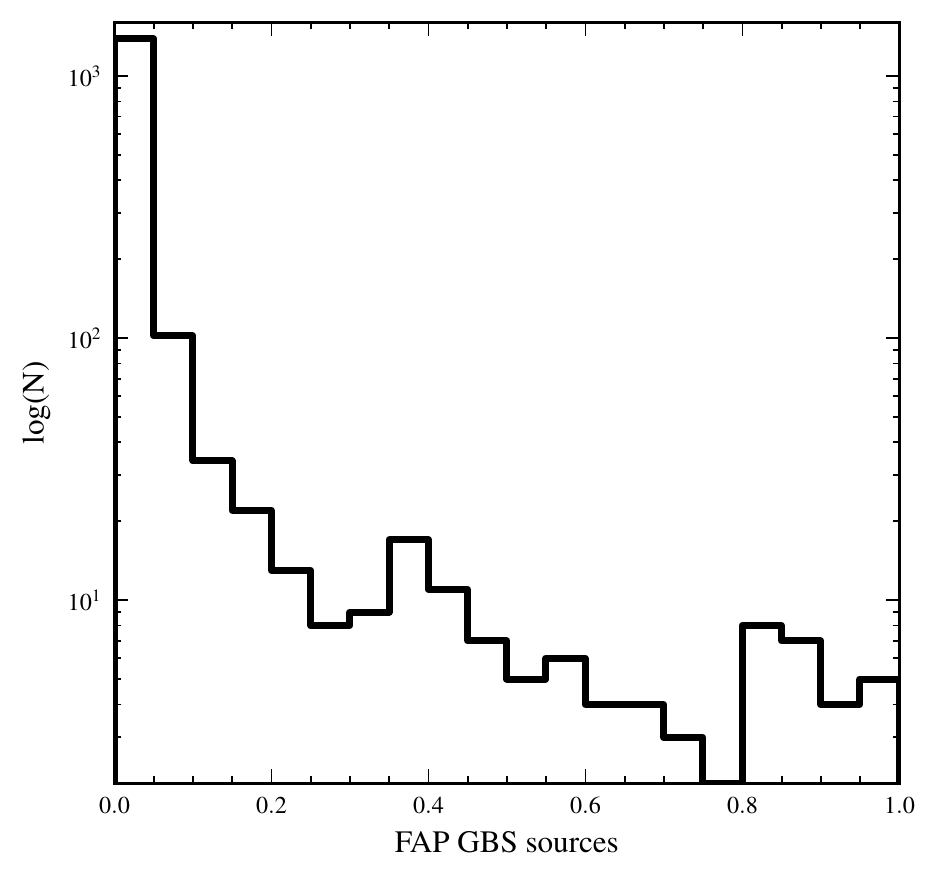}
\caption{Total FAP distribution of the GBS sources. Around 90\% of the sources have a final FAP $<$ 0.1 and $\sim$79\% have FAP $<$ 0.03.\label{fap-gbs}}
\end{figure}

We assign a final FAP ($FAP_\mathrm{final}$) to each viable matched source by taking into account the FAP calculated through the cross-matching of random sources in the GBS area with VVV and the FAP based on the positional uncertainties of the GBS sources: $FAP_\mathrm{final} = FAP_\mathrm{position} \times FAP_\mathrm{random}$. We show the distribution of $FAP_\mathrm{final}$ in Fig.\,\ref{fap-gbs}, where $\sim$90\% (1490 sources) of the GBS sources have a final FAP $<$ 10\% and $\sim$79\% of them have a final FAP $<$ 3\%. Even though 10\% remains a high value in terms of false alarm probabilities, it nonetheless confirms that we are not dominated by false matches to field stars and that the NIR matches found for most of the GBS sources are genuine counterpart candidates. About $\sim$3\% of the sources in the VVV NIR $K_s$ band catalogue did not have a valid $K_s$-band magnitude within R$_\mathrm{95}$, so they could not have a final FAP assigned to them.

\subsection{Multiple matches}

\begin{figure}
\hspace{-0.5cm}
\includegraphics[trim= 0cm 0cm 0cm 0cm, clip, scale=0.95]{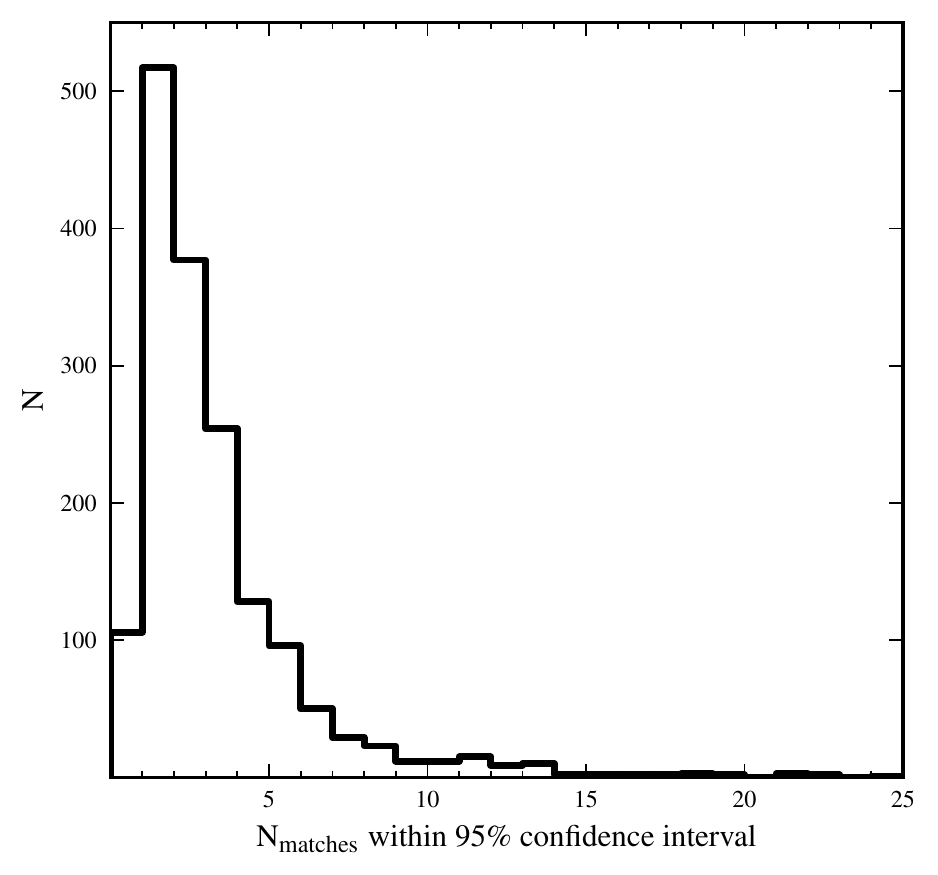}
\caption{Number of VVV matches found in a 95\% confidence interval (R$_\mathrm{95}$) from each X-ray source. The median value of this distribution is 2, meaning that each GBS source had typically 2 potential NIR matches in its R$_\mathrm{95}$ positional error radius.\label{n-3sig}}
\end{figure}

Our $FAP_\mathrm{final} = FAP_\mathrm{position} \times FAP_\mathrm{random}$ combines the fact that for larger source distances, FAP rate are higher due to field star contamination, but at the same time the probability that these are genuine matches is reduced.  More than one match may be consistent with our GBS source position and the closest match is not necessarily the best match. In order to identify the most likely counterpart to the X-ray source (i.e. the match with the lowest $FAP_\mathrm{final}$), we repeat the same process of calculating $FAP_\mathrm{position}$ and $FAP_\mathrm{random}$ for all the NIR matches within R$_\mathrm{95}$ of the total positional uncertainty of the source. We show in Fig.\,\ref{n-3sig} that typically, the GBS sources have 2 possible matches within R$_\mathrm{95}$. This value comes from the median of the distribution shown in Fig.\,\ref{n-3sig}.

\begin{figure}
\hspace{-1cm}
\includegraphics[trim= 0.cm 0cm 0cm 0cm, clip, scale=1]{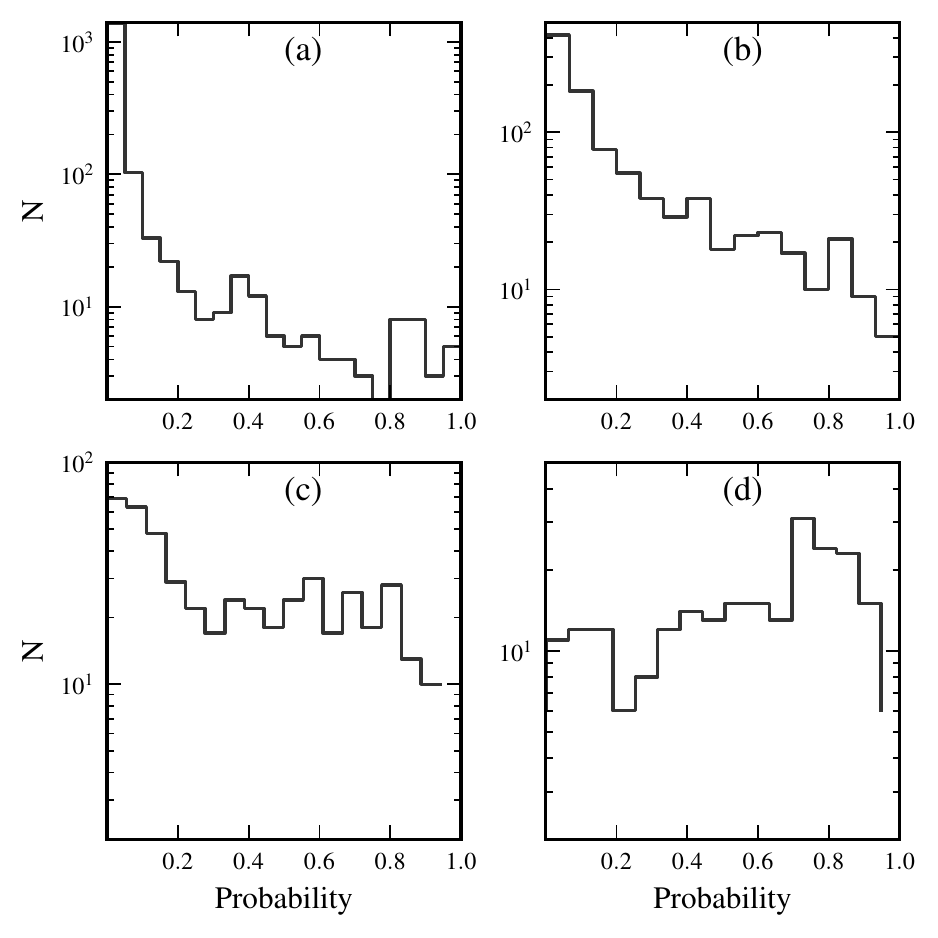}
\caption{$FAP_\mathrm{final}$ of four closest matches, within R$_\mathrm{95}$ of the X-ray position. Panels (a), (b), (c) and (d) correspond to the distributions of $FAP_\mathrm{final}$ of the closest, the second closest, the third closest and the fourth closest matches to the X-ray position. The total number of sources clearly drops as we move away from the X-ray position. \label{fap-matches}}
\end{figure}

In Fig.\,\ref{fap-matches}, we show the $FAP_\mathrm{final}$ for the nearest VVV matches (panel $a$), as well as the second (panel $b$), third (panel $c$) and fourth (panel $d$) closest matches within R$_\mathrm{95}$. We clearly see that the $FAP_\mathrm{final}$ increases as we move further away from the X-ray position. This indicates that the closest match has the most likely chance of being a real match, since the fourth closest VVV match has a typical $FAP_\mathrm{final}$ of 80\%. In addition, the number of sources with a fourth match within R$_\mathrm{95}$ decreases. Even though it is clear that the closest match is most likely to be the one with the lowest $FAP_\mathrm{final}$, we found that in 50 cases, the second closest match had a slightly lower $FAP_\mathrm{final}$ than the nearest one. This only represents $\sim$3\% of the sources but it is important to note. In such cases, the distances between the closest and second closest matches are similar yet the second closest match is brighter than the nearest one, making it a statistically more likely real counterpart to the X-ray source.

\begin{figure}
\centering
\includegraphics[trim= 0cm 0cm 0cm 0cm, clip, scale=0.95]{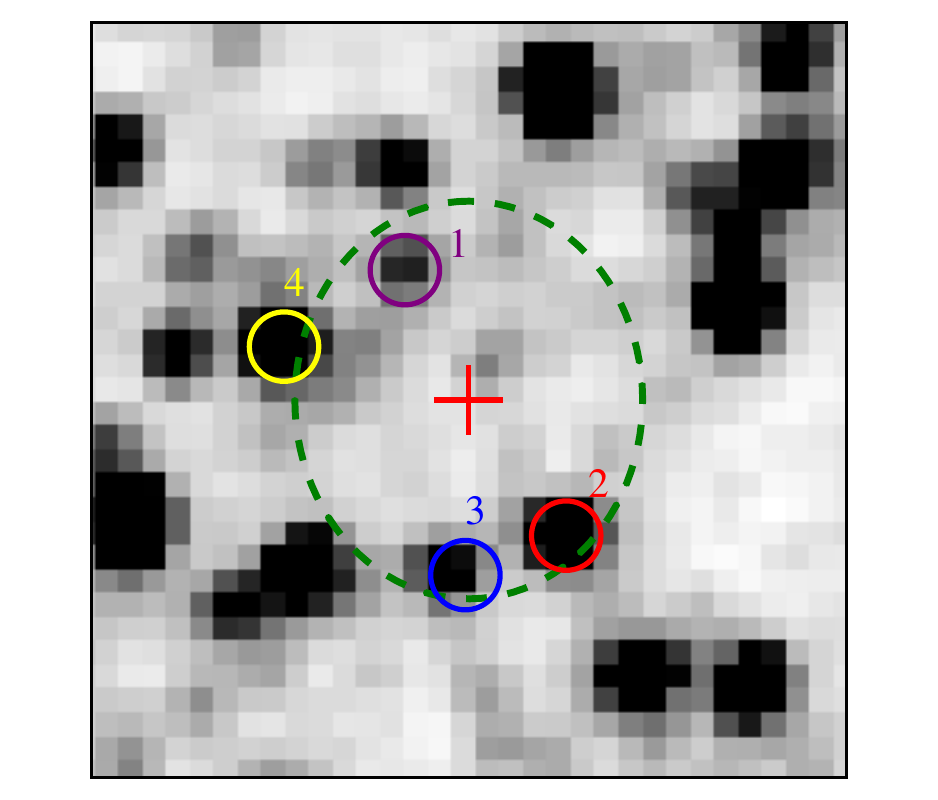}
\caption{Positions of the 4 closest matches of CX0013 found within R$_\mathrm{95}$ in VVV. The red cross indicates the X-ray position and the large dashed green circle indicates the R$_\mathrm{95}$ boundary of 2.84 arcseconds in this case. The table below provides information on their magnitudes and false alarm probabilities. \label{CX0013}}
\end{figure}

To illustrate this, we consider CX0013 as an example. In this case, we find 4 matches within its R$_\mathrm{95}$ (see Fig.\,\ref{CX0013} and Table\,\ref{CX0013-table}). As we move further away from the X-ray position, the final $FAP$ does increase dramatically making the closest match the preferred choice. However, upon inspection of the images, we notice a very faint object even closer to the X-ray position. This source is too faint to make it into the NIR source catalogues considered in this study. This is an important reminder that despite our analysis, we must always consider the possibility of even fainter sources not detected in VVV. Our source table identifies the most likely counterpart among the {\it detected} sources in VVV, 2MASS and UKIDSS GPS. 


\subsection{Final table}
\label{final-table}

To assist the characterization of the GBS source population, we provide in Table\,\ref{big-table} the NIR positions, magnitudes and calculated $K_s$-band FAP$_{final}$ for all the detected sources within R$_\mathrm{95}$ of the GBS X-ray positions. Such a Table presents a useful resource to anyone interested in studying the GBS sources, in particular at longer wavelengths and can be downloaded through the web version of this article. Here we only show the results for the first 30 brightest sources as an example of the full table, which contains 4661 entries.

\subsection{Influence of the hardness of the X-ray sources}

Since for most GBS sources, only a few counts are detected across the full 0.3 to 8 keV energy band, we have hardness ratios for the 164 brightest X-ray sources in GBS only. For the remaining GBS object, we can consider the energy range over which the sources were detected in. Therefore they can be given, when available, a hardness classification: {\it soft} X-ray sources are detected in the 0.3 to 2.5 keV band, while {\it hard} X-ray sources are detected solely in the 2.5 to 8 keV band.\\

\begin{figure*}
\centering
\includegraphics[trim= 0cm 0cm 0cm 0cm,  clip,  scale=0.9]{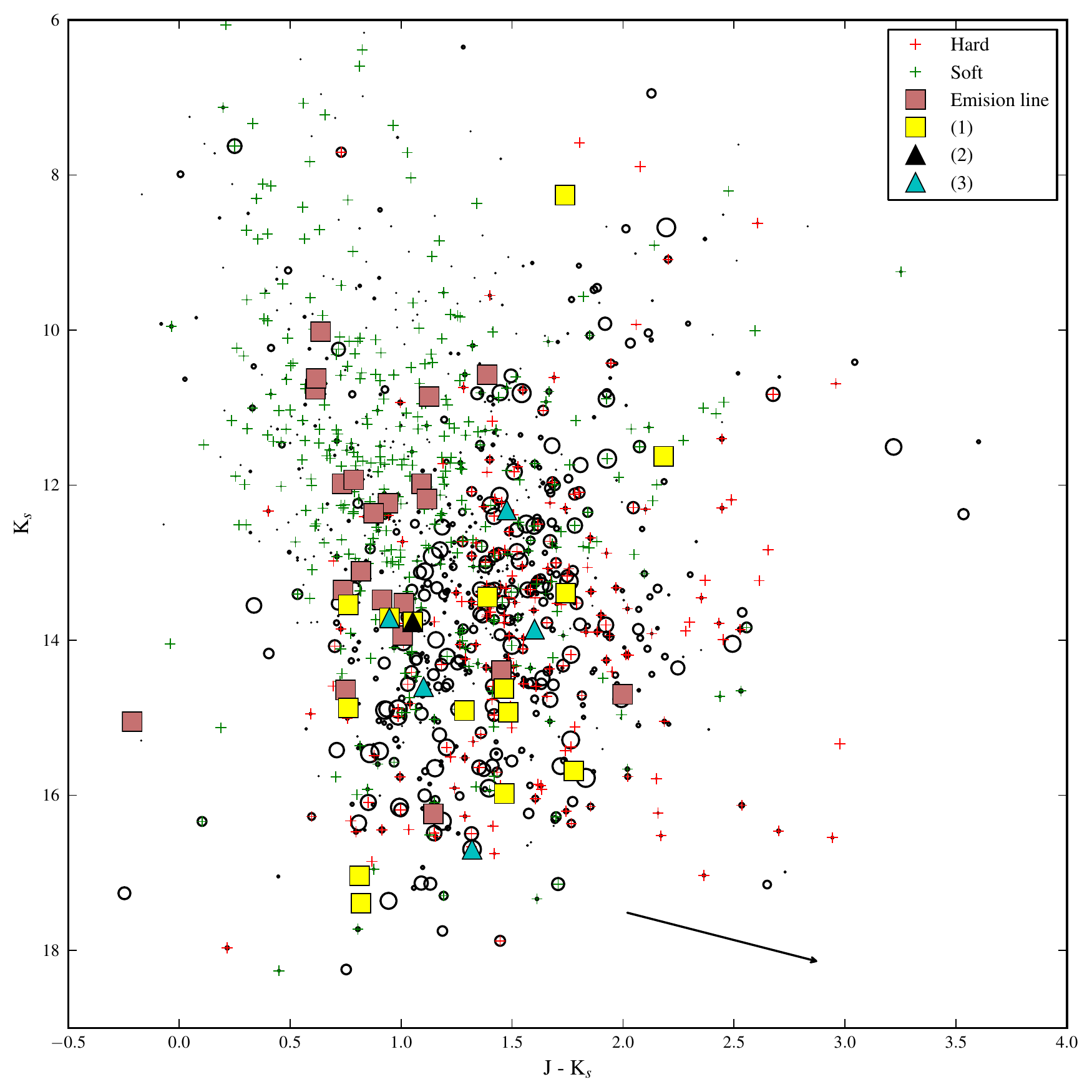}
\caption{($K_s$) vs ($J-K_s$) colour-magnitude diagram of the VVV matches with $FAP_\mathrm{final}$ $<$ 0.1. The size of the circle is proportionate to the value of the $FAP_\mathrm{final}$ of the source. The larger the circle, the bigger the $FAP_\mathrm{final}$. The red and green crosses correspond to the hard and soft X-ray sources respectively. The pink squares are H$\alpha$ emission line sources (AGN, M-stars, RS CVns) and the yellow squares are accreting binaries \citep{torresetal13}, all confirmed via spectroscopy. The black triangle corresponds to CX0093, a CV confirmed by \citet{rattietal13} and the cyan triangles correspond to the CVs studied in \citet{brittetal13}. The black arrow indicates the reddening for an extinction value of E(B~-~V)~=~1.8. \label{cmd-emitters3}}
\end{figure*}

We find that 327 sources are detected in the {\it soft} band, 444 are {\it hard} and the rest do not have a classification. In order to see if there is a correlation between the hardness of the X-ray sources, their NIR colours and the reddening towards the GBS fields, we plot a ($J - K_s$,~$K_s$) colour-magnitude diagram of the closest VVV matches to the GBS sources (see Fig.\,\ref{cmd-emitters3}). These sources are then colour-coded according to the X-ray hardness of the X-ray source, where red and green crosses correspond to {\it hard} and {\it soft} sources respectively. We also add a reddening vector with E(B-V)~=~1.8, since it corresponds to the typical extinction value towards the GBS region. Looking at the colour-magnitude diagram, we notice that {\it soft} sources are less affected by reddening than hard X-ray sources. This is an indication that the {\it soft} X-ray sources are more likely to be foreground objects whereas {\it hard} X-ray sources are probably reddened sources lying behind significant layers of extinction. Also most {\it soft} sources seem to have bright NIR matches, making them foreground sources and most probably the real matches. In order to confirm this result, we look at the $FAP_\mathrm{final}$ values of the {\it soft} and {\it hard} sources. We find that 90\% of the {\it soft} sources have a $FAP_\mathrm{final} < $~3\%, whereas this is the case for only 68\% of the {\it hard} sources. This indicates that we have probably found the real (foreground) NIR counterparts to the {\it soft} GBS sources.  \\

\begin{table}
\caption{Five closest VVV matches to CX0013. \label{CX0013-table}}
\centering
\begin{tabular}{c c c c}
\hline
Source & Distance & $K_s$ & $FAP_\mathrm{final}$ \\
 \hline \hline
1 & 2.07 & 15.60 &  0.069 \\
2 & 2.40 & 14.36 &  0.138 \\
3 & 2.50 & 15.38 &  0.188 \\
4 & 2.75 & 14.02 &  0.349 \\
\end{tabular}
\end{table}

\section{Discussion}
\label{discussion}

The combination of the NIR colours with the hardness of X-ray sources can provide additional information on the nature of the object beyond its proximity to the X-ray source. However, given the significant FAP for even the best matches, additional data is desired. Indeed an optical component is a key part of the GBS strategy. This consists of both optical imaging as well as spectroscopy. Here we briefly compare the NIR colour of some of these confirmed H$\alpha$ emission line objects.

\subsection{NIR colours of H$\alpha$ emission line objects}

In Fig.\,\ref{cmd-emitters3}, we show in pink squares some of the confirmed H$\alpha$ emission line sources in the GBS catalogue such as AGN, single M-stars and RS CVn systems. Also, \citet{torresetal13} have obtained spectra for several GBS sources and 23 objects show H$\alpha$ emission in their spectra, as well as accretion signatures. These types of X-ray binaries are our principle science targets and are plotted in yellow squares. As can be seen, many H$\alpha$ emission line objects fall in the region populated by the soft X-ray sources, possibly indicating that they are not Bulge sources. We also notice that most X-ray binary systems occupy the same region as the hard X-ray sources. However, they do not occupy a very distinct region of the diagram, making the source classification difficult when using NIR colours alone. The black triangle corresponds to CX0093 (also known as CX0153), a CV studied by \citet{rattietal13}, and the cyan triangles are the CVs published by \citet{brittetal13}.\\

\subsection{Towards the identification of key GBS source classes}

Despite the various studies of Galactic Centre X-ray sources using NIR photometry and spectroscopy (see Section\,\ref{intro}), classifying objects on the basis on their colours alone is difficult. The Galactic Centre and Bulge suffer from different amounts of extinction, which in itself varies on very small scales, greatly altering the colours. Therefore, strategies driven by a source's position in a colour-colour or colour-magnitude diagram, cannot be directly employed in these environments even though such methods are highly effective at high Galactic latitudes. 

In addition to the effects of reddening, the intrinsic colours of the sources expected in the GBS show great diversity. The key source types include LMXBs, CVs, UCXBs, RS CVn stars, W UMa and Be X-ray binaries. In the case of quiescent LMXBs, one may expect the companion stars to dominate the SED in the NIR. As these are typically late-type dwarfs, such objects may indeed have colours very similar to reddened field dwarfs. Comparison with theoretical colours of main-sequence and giant stars, as well as the correct reddening towards the line of sight of the X-ray source, can then help with the identification of potential NIR counterparts to the GBS sources. However, the presence of accretion continuum sources such as from accretion discs and jets will alter the colours. This diversity even among one subclass of, for example LMXBS, means that the NIR colours need complimentary constraints from other wavelength studies for a reliable classification of sources. 

Previous work suggests that the NIR colours of CVs are similar to F-K main-sequence stars \citep{hoardetal02}. Since most CVs are foreground objects, they do not suffer from the same amount of reddening as potential Bulge LMXBs. Note that this does not mean that their donor stars have spectral types of F-K as also in CVs accretion components will contribute. \citet{kniggeetal11} also found that NIR colours of CVs were dominated by the donor star, except for systems close to the period minimum where contributions from the WD were beginning to be more significant. Indeed, the majority of these systems have optical spectra dominated by the WD \citep{gansickeetal09}. Given their very low mass donors, the NIR colours of such systems thus no longer track a simple donor star sequence. It is also important to note that $\sim$ 20\% of CVs contain a magnetic WD (polars or intermediate polars). It has been found that they contribute towards a large fraction of the hard X-ray sources in the Galactic Centre \citep{munoetal04, hong12, brittetal13} since their X-ray luminosities are significantly higher than that of non-magnetic CVs. Luminosity ratios such as $L_{opt}/L_{X}$ can often be used as a crude discriminant between some of these source classes that otherwise may have similar NIR colours. Due to the fact the GBS is a shallow survey, with 2ks exposures, most detected sources have typically less than ten X-ray counts, leading to very poorly constrained X-ray fluxes. Another contribution to the large uncertainty of F$_X$ comes from the fact that the reddening towards the GBS X-ray sources is unknown. The VVV extinction maps yield a maximum limit to A$_{Ks}$, making it difficult to determine the actual X-ray flux of our sources. For this reason, we are unable to calculate reliable F$_X$/F$_{NIR}$ in order to identify key GBS source classes.

RS CVn stars, a type of close detached binary stars, are known to be variable due to cool stellar spots present on the surfaces. A good way to select them is by exploiting the NIR variability information which is now available in VVV and which will be the main topic of a future paper. 

The work presented here allows us to prioritize those with lowest $FAP_\mathrm{final}$ for spectroscopic or photometric follow-up and also assess the impact of false matches. The false alarm probability study is most reliable in the near-infrared (mainly in the $K_s$-band) since we can probe through the dust and detect more Bulge sources. With the final table presented in this study, which contains the most likely NIR counterparts to the GBS X-ray sources, we are now able to move on to the next stage of the GBS strategy, which is to use optical photometric and variability data to select the objects for spectroscopic follow-up. This has been demonstrated with the results found by \cite{torresetal13}, where key GBS source classes have been identified via spectroscopy. The addition of optical data will enable us to disentangle the effects of reddening towards the GBS fields and separate the field CVs from the Bulge LMXBs. The NIR variability information provided via VVV in the $K_s$-band will also help us in selecting those viable counterparts that show evidence for variability, as would be expected for the majority of our objects. True secure classification, however, is best achieved through spectroscopy (\citealt{torresetal13}, Wu et al. in prep).

\section{Conclusion}

We exploited three NIR surveys of the Galactic Bulge to search for the NIR counterparts of the GBS X-ray sources. We found that VVV was the most uniform survey, in terms of coverage and depth. We exploit the NIR data, along with the X-ray information, in order to find the NIR counterparts of the {\it Chandra} sources. We quantify the false alarm rate of finding the real matches by calculating false alarm probabilities for each source, taking into account their NIR magnitudes, distances to their matches, positional uncertainties and the multiple matches around each GBS object. We present these findings here in the form of a large data table (see Table\,\ref{big-table} for a subset of the final version available online) that will be a useful resource for follow-up studies of the GBS sources. We find that $\sim$90\% of the GBS sources have a $FAP_\mathrm{final}$ $<$ 10\% and $\sim$79\% of them have a $FAP_\mathrm{final}$ $<$ 3\%. This indicates that we have found the NIR counterpart of more than half of the GBS sources. We have shown that there are typically 2 NIR matches within R$_\mathrm{95}$ of the X-ray position but at least one of those matches is very likely to be the real counterpart. \\

While spectroscopy ultimately is a superior way of classifying key sources, such as the Bulge population of X-ray binaries, the ability to select candidates by using astronomical surveys and their photometric properties is crucial \citep{hynes10, motchetal09}. The NIR photometric data discussed here are able to probe through the dust and surveys such as VVV now have the spatial resolution to resolve these environments adequately. Our results can now be used in concert with data of the GBS region at other wavelengths, in order to disentangle the effects of reddening. This will lead to a more tailored target follow-up strategy. 

\section*{Acknowledgments}
SG acknowledges support through a Warwick Postgraduate Research Scholarship. DS acknowledges support from STFC through an Advanced Fellowship (PP/D005914/1) as well as grant ST/I001719/1. RIH and CTB acknowledge support from the National Science Foundation under Grant No. AST-0908789.
We gratefully acknowledge use of data from the ESO Public Survey programme ID 179.B-2002 taken with the VISTA telescope and data products from CASU.
We warmly thank D. Minniti and P. Lucas for early access to the VVV data, as well as O. Gonzalez early access to his calculated reddening values. We are grateful for all the help provided by R. Saito and M. Irwin with the access and reduction of the CASU data.


\landscape
\begin{table}
\caption{Table containing all NIR data and FAP of matches within R$_{95}$} \label{big-table}
\begin{tabular}{c c c c c c c c c c c c c c c}
GBS source & RA GBS &  Dec GBS & RA NIR & Dec NIR & Offset & J mag & J err & H mag & H err & K mag & K err & Survey & $FAP_\mathrm{final}$ & Comments\\
\hline
1 & 17 50 24.44 & -29 02 16.4 & 17 50 24.55 & -29 02 15.6 & 1.793 & 15.908 & 0.027 & 13.284 & 0.028 & 12.374 & 0.033 & 2MASS  & 3.161e-02 &  -  \\
2 & 17 37 28.39 & -29 08 02.0 & 17 37 28.39 & -29 08 02.1 & 0.034 & 13.633 & 0.056 & 12.257 & 0.054 & 11.187 & 0.059 &  2MASS  & 2.851e-09 &  (a)  \\
3 & 17 40 42.81 & -28 18 08.0 & 17 40 42.96 & -28 18 11.5 & 3.998 & 9.075 & 0.214 & 7.598 & 0.212 & 6.947 & 0.254 &  2MASS  & 2.176e-02 &  -  \\
4 & 17 39 31.22 & -29 09 52.8 & 17 39 31.22 & -29 09 53.3 & 0.515 & 7.209 & 0.001 & 6.566 & 0.003 & 6.384 & 0.004 &   2MASS  & 1.369e-08 &  (b)  \\
5 & 17 40 09.13 & -28 47 25.6 & 17 40 09.21 & -28 47 25.9 & 0.977 & 15.459 & 0.047 & 14.451 & 0.018 & 13.858 & 0.02 &   VVV  & 1.759e-03 &  (c)  \\
6 & 17 44 45.78 & -27 13 44.5 & 17 44 45.77 & -27 13 44.4 & 0.112 & 7.054 & 0.057 & 6.843 & 0.051 & 6.507 & 0.055 &   2MASS  & 8.455e-11 &  (b) \\ 
6 & 17 44 45.78 & -27 13 44.5 & 17 44 45.85 & -27 13 45.0 & 1.109 &  -  &  -  &  -  &  -  & 10.501 & 0.002 &   VVV  & 1.548e-03 &  -  \\
6 & 17 44 45.78 & -27 13 44.5 & 17 44 45.70 & -27 13 45.1 & 1.302 & 10.927 & 0.001 &  -  &  -  & 10.46 & 0.002 &   VVV  & 3.104e-03 &  - \\  
7 & 17 38 26.18 & -29 01 49.4 & 17 38 26.21 & -29 01 49.5 & 0.323 & 9.388 & 0.006 & 8.916 & 0.011 & 8.823 & 0.017 &   2MASS  & 6.359e-09 &  (b)  \\
8 & 17 35 08.28 & -29 29 57.9 & 17 35 08.24 & -29 29 58.2 & 0.556 &  -  &  -  &  -  &  -  & 10.665 & 0.002 &   VVV  & 7.200e-06 &  -  \\
9 & 17 35 08.40 & -29 23 28.4 & 17 35 08.42 & -29 23 28.3 & 0.169 & 10.162 & 0.006 & 9.828 & - & 9.659 & 0.006 &   2MASS  & 4.857e-09 &  (b)  \\
10 & 17 36 29.04 & -29 10 28.8 & 17 36 29.06 & -29 10 29.1 & 0.470 & 7.967 & 0.019 & 7.464 & 0.021 & 7.264 & 0.025 &   2MASS  & 3.453e-07 &  (b)  \\
11 & 17 41 51.30 & -27 02 23.5 & 17 41 51.42 & -27 02 23.8 & 1.705 & 15.146 & 0.014 & 14.04 & 0.012 & 13.505 & 0.013 &   VVV  & 1.957e-02 &  -  \\
12 & 17 43 47.24 & -31 40 25.2 & 17 43 47.26 & -31 40 25.2 & 0.411 & 6.998 & 0.001 & 6.364 & 0.001 & 6.165 & 0.001 &   2MASS  & 3.020e-08 &  (b) \\ 
13 & 17 50 29.13 & -29 00 02.3 & 17 50 29.20 & -29 00 00.5 & 2.066 &  -  &  -  &  -  &  -  & 15.599 & - &   VVV  & 6.935e-02 &  -  \\
13 & 17 50 29.13 & -29 00 02.3 & 17 50 29.03 & -29 00 04.3 & 2.402 & 16.587 & 0.11 & 14.92 & 0.076 & 14.363 & 0.074 &   VVV  & 1.380e-01 &  -  \\
13 & 17 50 29.13 & -29 00 02.3 & 17 50 29.14 & -29 00 04.8 & 2.500 & 17.253 & 0.194 &  -  &  -  & 15.376 & 0.179 &   VVV  & 1.880e-01 &  -  \\
13 & 17 50 29.13 & -29 00 02.3 & 17 50 29.33 & -29 00 01.6 & 2.745 & 16.203 & 0.078 & 14.58 & 0.056 & 14.02 & 0.055 &   VVV  & 3.349e-01 &  -  \\
14 & 17 46 23.67 & -31 35 00.8 & 17 46 23.69 & -31 35 00.6 & 0.191 & 9.962 & 0.33 & 9.262 & - & 9.119 & - &   2MASS  & 1.355e-09 &  -  \\
15 & 17 46 46.17 & -25 52 17.5 & 17 46 46.25 & -25 52 17.5 & 0.948 & 16.198 & 0.23 & 15.571 & 0.409 & 15.392 & - &   VVV  & 1.381e-03 &  -  \\
16 & 17 55 45.83 & -27 58 14.0 & 17 55 45.84 & -27 58 13.8 & 0.288 & 12.908 & - & 12.104 & - & 11.72 & 0.499 &   2MASS  & 3.330e-07 &  -  \\
16 & 17 55 45.83 & -27 58 14.0 & 17 55 45.84 & -27 58 15.8 & 1.843 & 15.269 & 0.017 &  -  &  -  & 14.036 & 0.036 &   VVV  & 4.314e-02 &  -  \\
17 & 17 52 53.02 & -29 22 09.1 & 17 52 52.96 & -29 22 08.0 & 1.383 &  -  &  -  & 13.778 & - & 13.538 & - &   VVV  & 7.763e-03 &  -  \\
17 & 17 52 53.02 & -29 22 09.1 & 17 52 53.14 & -29 22 09.4 & 1.631 & 12.437 & 0.036 & 11.27 & 0.031 & 10.893 & 0.069 &   2MASS  & 1.124e-02 &  -  \\
18 & 17 39 35.77 & -27 29 35.9 & 17 39 35.76 & -27 29 36.0 & 0.142 & 15.702 & 0.001 & 14.93 & 0.001 & 14.601 & 0.001 &   VVV  & 2.422e-08 &  (c)  \\
19 & 17 49 54.57 & -29 43 35.4 & 17 49 54.53 & -29 43 35.8 & 0.745 & 11.989 & - & 10.432 & - & 9.93 & 0.046 &   2MASS  & 5.201e-05 &  -  \\
20 & 17 38 59.68 & -28 24 49.1 & 17 38 59.66 & -28 24 49.5 & 0.407 & 15.768 & 0.018 & 15.117 & 0.015 & 14.636 & 0.016 &   VVV  & 7.970e-06 &  -  \\
21 & 17 41 33.76 & -28 40 33.8 & 17 41 33.79 & -28 40 34.5 & 0.890 & 17.284 & 0.093 & 16.849 & 0.055 & 16.447 & 0.058 &   UKIDSS  & 8.253e-04 &  - \\ 
22 & 17 45 54.02 & -31 15 03.4 & 17 45 54.01 & -31 15 03.2 & 0.146 & 10.27 & - & 9.619 & - & 9.472 & - &   2MASS  & 1.725e-09 &  -  \\
23 & 17 42 31.56 & -27 43 49.1 & 17 42 31.56 & -27 43 48.3 & 0.808 & 11.048 & 0.065 & 9.527 & 0.056 & 8.907 & 0.053 &   2MASS  & 7.205e-05 &  -  \\
24 & 17 48 49.51 & -30 01 09.9 & 17 48 49.50 & -30 01 10.3 & 0.368 & 10.947 & 0.001 & 10.217 & 0.002 & 9.965 & 0.002 &   2MASS  & 3.882e-08 &  -  \\
24 & 17 48 49.51 & -30 01 09.9 & 17 48 49.71 & -30 01 09.7 & 2.559 &  -  &  -  & 13.864 & 0.019 & 13.363 & 0.024 &   VVV  & 2.020e-01 &  -  \\
25 & 17 45 02.78 & -31 59 35.0 & 17 45 02.77 & -31 59 34.5 & 0.704 & 10.175 & 0.023 & 9.719 & 0.039 & 9.584 & 0.059 &   2MASS  & 3.018e-05 &  (b) \\ 
26 & 17 45 33.26 & -30 58 56.2 & 17 45 33.27 & -30 58 55.9 & 0.122 & 7.666 & 0.017 & 7.395 & 0.017 & 7.335 & 0.023 &   2MASS  & 1.069e-10 &  (b)  \\
27 & 17 36 52.83 & -28 48 41.6 & 17 36 52.83 & -28 48 41.4 & 0.130 & 11.621 & 0.022 &  -  &  -  & 10.524 & 0.03 &   VVV  & 2.664e-10 &  (b)-Saturated  \\
28 & 17 39 46.98 & -27 18 09.5 & 17 39 47.01 & -27 18 08.9 & 0.680 & 14.658 & 0.002 & 14.009 & 0.002 & 13.711 & 0.002 &   VVV  & 1.213e-04 &  (c)  \\
29 & 17 53 41.91 & -28 03 53.4 & 17 53 41.89 & -28 03 53.8 & 0.373 & 13.654 & 0.168 & 12.773 & 0.3 & 12.502 & 0.345 &   2MASS  & 2.673e-06 &  -  \\
30 & 17 49 20.62 & -30 18 31.8 & 17 49 20.55 & -30 18 32.4 & 1.045 & 16.618 & 0.039 & 15.697 & - & 15.369 & - &   VVV  & 2.219e-03 &  -  \\
30 & 17 49 20.62 & -30 18 31.8 & 17 49 20.68 & -30 18 31.0 & 1.101 & 17.861 & 0.164 & 16.542 & 0.196 & 16.266 & 0.296 &   VVV  & 3.029e-03 &  -  \\
\hline
\end{tabular}
\\ 
(a) \cite{maccaroneetal12}, (b) \cite{hynesetal12}, (c) \cite{brittetal13}, (d) \cite{rattietal13}
\end{table}
\endlandscape

\label{lastpage}

\end{document}